\documentclass[a4paper,11pt]{article}

\pdfoutput=1 

\usepackage{jheppub} 

\usepackage[T1]{fontenc} 

\usepackage{booktabs}

\usepackage{etex,etoolbox}
\usepackage{amsthm}

\chardef\MyArticleWithColor=\pdfcolorstackinit page direct{0 g}

\newcommand\mgamc{{\sc\small MG5aMC}}
\newcommand\madgraphamc{{\sc\small MadGraph5\_aMC@NLO}}

\newcommand\MadLoop{{\sc\small MadLoop}}
\newcommand\MadEvent{{\sc\small MadEvent}}

\newcommand\madevent{{\sc\small MadEvent}}
\newcommand\feynrules{{\sc\small FeynRules}}
\newcommand\pythia{{\sc\small Pythia}}
\newcommand\delphes{{\sc\small Delphes}}
\newcommand\ufo{{\sc\small UFO}}

\def\d#1{D_{#1}}
\def\db#1{\bar D_{#1}}

\def\beq{\begin{equation}}
\def\eeq{\end{equation}}
\def\beqn{\begin{eqnarray}}
\def\eeqn{\end{eqnarray}}

\newcommand\CutTools{{\sc\small CutTools}}
\newcommand\MadGraph{{\sc\small MadGraph}}

\newcommand\syscalc{{\sc\small SysCalc}}

\newcommand\Sherpa{{\sc\small SHERPA}}

\newcommand\madspin{{\sc\small MadSpin}}

\newcommand\MLf{{\sc\small MadLoop5}}

\newcommand\qloop{\ell}
\newcommand\bqloop{\bar{\ell}}

\title{Automated event generation for loop-induced processes}
\preprint{IPPP/15/35, DCPT/15/70, MCNET/15/14, SLAC-PUB-16369}

\author[a]{Valentin Hirschi}
\author[b]{Olivier Mattelaer}

\affiliation[a]{SLAC National Accelerator Laboratory, 2575 Sand Hill Road, Menlo Park, CA 94025-7090 USA}
\affiliation[b]{Institute  for  Particle  Physics  Phenomenology  (IPPP),
Durham  University,  Durham,  DH1  3LF,  United  Kingdom}

\emailAdd{vahirsch@slac.stanford.edu}
\emailAdd{o.p.c.mattelaer@durham.ac.uk}

\abstract{
We present the first fully automated implementation of cross-section computation and event generation for loop-induced processes. This work is integrated in the \madgraphamc\ framework. We describe the optimisations implemented at the level of the matrix element evaluation, phase space integration and event generation allowing for the simulation of large multiplicity loop-induced processes. Along with some selected differential observables, we illustrate our results with a table showing inclusive cross-sections for \emph{all} loop-induced hadronic scattering processes with up to three final states in the SM as well as for some relevant $2\rightarrow 4$ processes. Many of these are computed here for the first time.}
\begin{document} 
\maketitle
\flushbottom

\section{Introduction}
\label{sec:intro}

The first run of the Large Hadron Collider (LHC) further confirmed the Standard Model (SM) in a spectacular way with the discovery of what was its last missing piece: the Higgs boson~\cite{Aad:2012tfa,Chatrchyan:2012ufa,Englert:1964et,Higgs:1964pj,Higgs:1964ia}. This discovery testifies not only of the technical success of the accelerator and detectors, but also of the level of accuracy reached in the theoretical predictions intervening at many levels of a discovery or exclusion at a collider. One can identify three main types of effort at the origin of this precision reached by modern high energy physics simulations. First, dedicated analytical single-purposed computations keep pushing the boundaries of perturbative physics and are essential for realistic inclusive predictions (\emph{e.g.} Higgs hadroproduction at N$^3$LO~\cite{Anastasiou:2015vya}). 
Secondly, parton shower Monte-Carlo programs such as {\sc Herwig++}~\cite{Bahr:2008pv}, {\sc Pythia8}~\cite{Sjostrand:2007gs} and {\sc Sherpa}~\cite{Gleisberg:2008ta} improved their formal control on the soft physics resummation as well as on the merging techniques with hard matrix-element predictions. Finally, the last decade has seen the rise of a number of automated one-loop matrix element computation tools, such as \MadLoop~\cite{Hirschi:2011pa}, {\sc \small FeynArts}~\cite{Hahn:2000kx,Hahn:1998yk}, {\sc \small OpenLoops}~\cite{Cascioli:2011va} and {\sc \small GoSam}~\cite{Cullen:2011ac}.
Thanks to these progresses, one can obtain accurate normalisation and distributions for basically any process of interest at NLO accuracy, using flexible partonic hard event generators such as {\sc \small POWHEG}~\cite{Alioli:2010xd,Frixione:2007vw,Nason:2004rx}, {\sc Sherpa} and the framework in which the present work is carried, \madgraphamc~\cite{Alwall:2014hca} (referred to as \mgamc\ henceforth).

The aforementioned tools are crucial for both experimentalists and model builders to simulate backgrounds and explore new physics signals. 
However, despite these efforts, event generation for \emph{loop-induced} processes, \emph{i.e.} processes without any tree-level contributions and starting at one-loop, is still not available in a systematic and fully automated way.
This is to be contrasted with the fact that loop-induced amplitudes for many relevant processes are already available in the main public one-loop matrix element providers~\cite{Hirschi:2011pa,Hahn:2000kx,Hahn:1998yk,Cascioli:2011va,Cullen:2011ac}.
The aim of this work is to remedy this situation by providing an efficient and generic technique for the simulation of loop-induced processes which play a significant role both in the SM and beyond. 
Loop-induced processes take their name from the fact that their leading-order (LO) contribution comes already from loop amplitudes, so that loop integrals are unavoidable in this case, even for the crudest computation of the cross-section. Exceptions to this are certain processes where the loop featuring heavy particles can be \emph{integrated out} to form an effective point-like vertex turning the loop topology into a tree one. However, this approximation is typically valid only in a limited kinematical range, so that the general implementation of the direct computation of loop-induced processes is desirable.

In the SM, a prime example is the gluon fusion channel for Higgs boson production which dominates the inclusive production, mainly because of the large gluon luminosity in high energy hadron colliders. Besides this obvious case, loop-induced channels can also amount to a significant part of some NNLO corrections; it was for example recently shown in ref.~\cite{Cascioli:2014yka} that the contribution of the loop-induced $g g \rightarrow Z Z$ channel represents 60\% of the full NNLO correction of hadronic Z-boson pair production.
Loop-induced processes are also often relevant in the context of beyond the Standard Model (BSM) models where the associated loop suppression factor can help to evade current experimental bounds.

The attitude towards the computation of loop-induced processes has been so far that of developing single-purpose specific codes (see references in sect. \ref{sec:results}). Given current loop technology, it is desirable to adopt a solution both generic in the process studied and flexible in its use. We present here the implementation of this solution within the public framework \mgamc. Loop matrix-elements are computed with \MadLoop, using a combination of the Ossola-Papadopoulos-Pittau (OPP)~\cite{Ossola:2008zza,ossola-2007-763} reduction method as implemented in \CutTools~\cite{Ossola:2007ax} and various Tensor Integral Reduction tools (TIR) such as {\sc \small PJFry}~\cite{Fleischer:2011bi}, {\sc \small Golem95}~\cite{Binoth:2008uq,Cullen:2010hz} and the in-house implementation   {\sc \small IREGI}. The integration over phase-space and event generation is performed by \MadEvent~\cite{Maltoni:2002qb}, the multi-purpose phase space integrator used for LO calculations in \mgamc.

The paper is organised as follows. 
Sect.~\ref{sec:method} describes the various improvements brought to \MadLoop\ and \MadEvent\ to cope with loop-induced processes, for which the absence of underlying tree topologies renders inapplicable many optimisations.
Sect.~\ref{sec:results} presents a comprehensive list of cross-sections for SM loop-induced processes obtained within our implementation. In sect.~\ref{sec:higgs}, we focus on loop-induced Higgs production to illustrate how the various simulation features of \mgamc\ apply in this case. A systematic comparison of our results with those obtained using the Effective Field Theory (EFT) approach shows where and to which extent the latter is a good approximation. Sect.~\ref{sec:Validation} is devoted to the validation of our implementation within a BSM context against the independent computation of ref.~\cite{Hespel:2015zea} for loop-induced Z-Higgs associated production. We also show results for the loop-induced charged Higgs pair production. We summarise our work in sect.~\ref{conclusion}.

\section{Technique}
\label{sec:method}
\subsection{Loop computation, TIR implementation and colour decomposition}
\label{techloop}

The main challenge for loop-induced processes integration is the running speed of the related loop-matrix element computation. In NLO computations, the bulk of the contribution comes from real-emission and Born topologies, so that only limited statistics is necessary for the evaluation of the virtual contribution (see sec. 2.4.3 of ref.~\cite{Alwall:2014hca}); it is then very often not the limiting factor in terms of computational load. The situation is radically different for the direct integration of loop-induced processes where the computation time is proportional to the execution speed of the loop matrix element.

To understand the characteristics of the various techniques available in \MadLoop\ for the computation of loop-induced squared matrix elements, it is appropriate to start from its generic expression:
\begin{eqnarray}
\left |\mathcal{A}^{LI}\right |^2 &=&  \left |\mathcal{A}^{LI}_{\text{non-$R_2$}}\right |^2+2\Re\left ( \mathcal{A}^{LI}_{\text{non-$R_2$}}\mathcal{A}^{LI*}_{\text{$R_2$}}\right )+\left | \mathcal{A}^{LI}_{\text{$R_2$}}\right |^2 \label{LIAmpDef}
\\
\nonumber
\left |\mathcal{A}^{LI}_{\text{non-$R_2$}}\right |^2 &=& \sum_{\rm colour}\sum_{h=1}^{H}
\left(\sum_{l_1=1}^{L} \lambda_{l_1}\int d^d \bqloop\,
\frac{ {\cal N}_{h,{l_1}}(\qloop)}{\prod_{i=1}^{n_{l_1}}\db{i,{l_1}}}\right)
\left(\sum_{l_2=1}^{L} \lambda_{l_2}\int d^d \bqloop\,
\frac{ {\cal N}_{h,{l_2}}(\qloop)}{\prod_{i=1}^{n_{l_2}}\db{i,{l_2}}}
\right)^\star,
\label{VBtemp}
\end{eqnarray}

where $\mathcal{A}^{LI}$ designates the loop-induced amplitude, $\lambda_{l_i}$ are colour structures, $\mathcal{N}_{h,l_i}$ are loop integrand numerators and $\db{i,l_j}$ are $d$-dimensional propagator denominators of the form $(\bqloop+k_{i,l_j})^2-m_{i,l_j}^2$. The symbols $\sum_h$, $\sum_{\text{colour}}$ and $\sum_{l_i}$ denote the sum over all helicity, colour configurations and loop subamplitudes factoring a single colour factor.\footnote{These subamplitudes are in one-to-one correspondence with the amplitudes of the constituting Feynman diagrams, except for those involving vertices featuring mlultiple colour factors (\emph{e.g.} four-gluon vertex).} The integers $H$, $L$ and $n_{l_i}$ are the total number of helicity configurations, loop subamplitudes and loop propagator denominators in subamplitude ${l_i}$ respectively. Notice that no UV counterterms are necessary in this case given that loop-induced processes are finite.
Conversely, R2 counterterms~\cite{Draggiotis:2009yb,Ossola:2008xq,Draggiotis:2009qy}, originating from the 4-dimensional nature of the reduction methods applied in \MadLoop, must be included.\footnote{Counterterms are selected and constructed with the same algorithm as for NLO virtual matrix element generation, \emph{i.e.} directly starting from the loop Feynman diagrams considered, hence guaranteeing the consistency of the computation.} The $|\mathcal{A}^{LI}_{\text{non-$R_2$}}|^2$ term of eq.~\ref{LIAmpDef} is the new element specific to loop-induced matrix element computations, and we now turn to detailing its implementation in \MadLoop.

The key characteristic of the quantity $|\mathcal{A}^{LI}_{\text{non-$R_2$}}|^2$ is that it is not linearly dependent on the loop amplitudes. This has dramatic implications on the type of optimisations applicable to its computation. When replacing the loop integral by the formal reduction operator $\text{Red[]}$ (symbolically denoting the application of any one of the available reduction tools interfaced to \MadLoop), one finds:
\begin{eqnarray}
\nonumber
\left |\mathcal{A}^{LI}_{\text{non-$R_2$}}\right |^2 &=&
\sum_{h=1}^{H} \sum_{l_1=1}^{L} \sum_{l_2=1}^{L} \left( 
\text{Red}\left [ \frac{{\cal N}_{h,{l_1}}(\qloop)}{\prod_{i=1}^{n_{l_1}}\db{i,{l_1}}} \right ]
\text{Red}\left [ \frac{{\cal N}_{h,{l_2}}(\qloop)}{\prod_{i=1}^{n_{l_2}}\db{i,{l_2}}} \right ]^* \underbrace{\sum_{\text{colour}}\lambda_{l_1}\lambda_{l_2}^*}_{\Lambda_{l_1,l_2}}
\right).
\end{eqnarray}
Contrary to NLO computations, where loop amplitudes interfere with tree-level ones, it is clear that the $\text{Red}[]$ operator cannot be pulled out of the sum over helicity configurations so as to apply at the squared matrix element level instead (see the transition from eq.~2.75 to eq.~2.76 of ref~\cite{Alwall:2014hca}). The expanded sum $\sum_{l_1=1}^{L} \sum_{l_2=1}^{L}$ contains $L^2$ terms and this quadratic scaling with the number of diagrams is problematic when compared to the apparent\footnote{This statement must be taken with care since both the construction and reduction of loops increases in complexity with the multiplicity. However, internal recycling of currents effectively implements recursive relations, mitigating this increased complexity and yielding a scaling approximately linear in the number of diagrams, at least for processes with up to four final state legs~\cite{Cascioli:2011va}.} linear scaling of the total loop computation time. As a result, the computation time becomes dominated by the squaring operation for loop-induced processes with as few as a thousand diagrams. To circumvent this scaling, we consider here the same solution as the one adopted for tree-level computations, namely colour decomposition.
Indeed, the origin of this problem can be traced back to the size of the colour basis, built out of a total of $L$ basis vectors $\lambda_{l}$ which are not linearly independent. The solution consists then in projecting the colour factors $\lambda_{l}$ onto a \emph{colour-flow} basis~\cite{Maltoni:2002mq} built out of $K$ colour-flow basis vectors $\kappa_i$ (chains of Kronecker delta structures with indices in the (anti-)fundamental representation of $SU(3)_c$). 
The growth of $K$ with the multiplicity is power-like, hence guaranteeing that the computational cost of the colour algebra involved in the amplitude squaring operation remains negligible with respect to that of the loop amplitude computation. For example, the process $g g \rightarrow h g g g$ has 3330 subamplitudes but only 24 different colour-flows. Each loop colour factor is then projected onto the colour-flow basis as follows:
\begin{equation}
\lambda_{l} = \sum_{i=1}^{K} \underbrace{(\lambda_{l}\otimes \kappa_i)}_{\alpha_{l,i}} \kappa_i.
\end{equation}
Notice that in practice, the projection matrix is sparse so that most of the projection coefficients $\alpha_{l,i}$ are zero and the sum involves a few terms only. The corresponding colour matrix is
\begin{equation}
\sum_{\text{colour}}\kappa_i \kappa_j^* = K_{ij}.
\end{equation}
Once expressed in the colour-flow basis, the loop-induced squared amplitude reads
\begin{eqnarray}
\left |\mathcal{A}^{LI}_{\text{non-$R_2$}}\right |^2 &=&
\sum_{h=1}^{H} \sum_{i=1}^{K} \sum_{j=1}^{K} \left(
J_{i,h}J_{j,h}^* K_{i,j}
\right)
\end{eqnarray}
where ${J_{i,h}}$ is defined as follows:
\begin{eqnarray}
J_{j,h}&:=&\sum_{l=1}^{L}\alpha_{i,l}\text{Red}\left [ \frac{{\cal N}_{{l},h}(\qloop)}{\prod_{i=1}^{n_{l}}\db{i,{l}}} \right ]
\end{eqnarray}
and corresponds to the partial colour amplitudes built from sums of the Lorentz part of the loop amplitudes weighted by projection coefficients.  This change of colour basis is not only computationally advantageous, but also more physical since the partial colour amplitudes are gauge invariant. It also automatically solves the problem of assigning a colour flow to partonic events, which amounts to specifying the colour dipole pairs in the starting conditions of parton shower Monte-Carlo programs and thus has an important impact on the radiation pattern.

This colour projection is performed automatically by \mgamc, using the same colour algebra module employed for tree-level matrix element generation. However, contrary to the tree-level case, the projection coefficients $\alpha_{l,i}$ can be of different orders in the colour expansion in $N_{c}$ because of the loop colour trace. This effect is accounted for when assigning colours to the generated events in which case only the leading term in the colour expansion is kept. 
Also, thanks to the flexibility of the colour module performing the algebra, arbitrary colour structures can be supported as well as the definition of any other basis, were that be necessary. The computation of the partial colour amplitudes has been made available also for the computation of the standard loop and Born interference term appearing in NLO computations (see appendix~\ref{subsec:newoptions}).

The inability to perform loop reduction at the squared matrix element level leads to a crucial difference between the TIR and OPP reduction methods. To understand why this is so, we detail the expression taken by the $\text{Red[]}$ operation in both cases. 
The integrand numerator is decomposed in a polynomial in the loop momentum $\ell$ as follows:
\beq
\label{eq:openloop}
\mathcal{N}(\ell)_{l,h}=\sum_{r=0}^{r_{max}} C^{(r)}_{\mu_1\ldots\mu_r;h,l} \;\qloop^{\mu_1}\ldots\qloop^{\mu_r}
\eeq
where $C^{(r)}_{\mu_1\ldots\mu_r;h,l}$ are referred to as the \emph{polynomial coefficients}.
In the case of OPP reduction, this decomposition only serves the purpose of improving the efficiency of the computation since the polynomial coefficients can be recycled for all the different values of the loop momentum for which $\mathcal{N}(\ell)_{l,h}$ must be evaluated. In the case of TIR, this decomposition is essential since the object reduced are the \emph{tensor integrals} $T^{(r)}_l$:
\beq
T^{(r),\mu_1 \cdots \mu_r}_l \equiv{\int d^d \bqloop\,\frac{\qloop^{\mu_1}\ldots\qloop^{\mu_r}}
{\prod_{i=1}^{n_{l}}\db{i,{l}}}}.
\eeq
We can now write the more precise form taken by OPP and TIR reduction:
\beq
\text{Red}\left [ \frac{{\cal N}_{{l},h}(\qloop)}{\prod_{i=0}^{n_{l}}\db{i,{l}}} \right ] =
\left \{ \begin{matrix}
\text{OPP}\left [ \frac{
\sum_{r=0}^{r_{max}} C^{(r)}_{\mu_1\ldots\mu_r;h,l} \;\qloop^{\mu_1}\ldots\qloop^{\mu_r}
}{\prod_{i=1}^{n_{l}}\db{i,{l}}} \right ] \\
\\
\sum_{r=0}^{r_{max}} C^{(r)}_{\mu_1\ldots\mu_r;h,l} \; \text{TIR}\;[ \; \frac{\qloop^{\mu_1}\ldots\qloop^{\mu_r}} {\prod_{i=1}^{n_{l}}\db{i,{l}}}\;]
\end{matrix}
\right. .
\eeq

It is now manifest that the output of the OPP reduction depends on both the loop and helicity considered while the TIR output only depends on the loop considered (\emph{i.e.} the tensor integrals $T^{(r)}_l$ only carry a dependence on the index $l$, not $h$). For this reason, the number of independent OPP reductions performed per kinematic configuration is necessarily $L\times H$, that is the number of loop amplitudes times the number of helicity configurations. On the other hand, the number of independent\footnote{They are not completely independent, since most TIR implementations internally cache some intermediate reduction results and scalar integral computations so that they can be re-used across loops sharing some of their reduced topologies.} TIR per phase-space points is only proportional to $L$ since tensor integrals can be recycled across helicity configurations.
For this reason, the evaluation of loop-induced matrix elements summed over helicity configurations is typically faster using TIR. Conversely, for the computation of a single helicity configuration, OPP type of reduction is preferable since the implementations of TIR currently available in \MadLoop, namely {\sc\small PJFry}, {\sc\small Golem95}\footnote{The current version of {\sc\small Golem95} does not currently allow tensorial coefficients to be recycled.} and the in-house implementation {\sc\small IREGI}, are slower for an equal number of calls to the $\text{Red[]}$ operator (see appendix~\ref{localRes} for quantitative results on benchmark processes). Note that this statement is highly dependent on the TIR implementation considered, and comparisons presented in ref.~\cite{Cascioli:2011va} suggest that it might not hold true for the private TIR implementation in {\sc\small COLLIER}~\cite{Denner:2010tr,Denner:2005nn}. However, due to the different scaling of the complexity of TIR and OPP reduction with the rank of the loop integral, it is expected that OPP is always faster for larger ranks (typically $r \gtrsim 6$).

Given the above, it is unclear whether event generation is more efficient using TIR and an explicit sum over helicity configurations or using OPP reduction and a Monte-Carlo (MC) sampling over them. The outcome mostly depends on the quality of the helicity discrete importance sampling ({i.e.} the magnitude of its mild dependence with kinematics), the runtime speed of TIR implementations and the relative importance of the computational cost of the determination of the polynomial coefficients (which scales like $L\times H$). Our tests show that generating events with an MC over helicity configurations, using an independent importance sampling for each integration channel, yields better timings and we use it as our default (see sect.~\ref{sec:PSint}).

We now turn to the description of two additional minor improvements on \MadLoop. First, many loop-induced processes occur only via closed fermion loops where often the different massless flavours bring the exact same contribution and can therefore be recycled. Identical diagrams (couplings, propagator spins, masses and widths are compared) are detected at generation level and then traded for a multiplicative factor affecting one chosen representative diagram.\footnote{This means that the loop diagrams removed this way no longer show in the list of diagrams drawn by \MadLoop.} For example, this leads to an improvement of a factor two for the production of electroweak bosons via gluon fusion. Secondly, because of Furry theorem, Feynman loop diagrams with an odd number of photon external legs can be exactly zero and their presence slows down the integration because \MadEvent\ must probe the corresponding channels many times before deciding that it can be safely discarded. To avoid this, \MadLoop\ has been modified so as to detect and remove such Furry loops at the diagram generation step (with a notification to the user).

The loop-induced matrix element codes generated by \MadLoop\ have been validated in several ways. First, we compared numerical results for specific kinematic configurations of many processes against \MadLoop4~\cite{Hirschi:2011pa}, whose implementation is completely independent of \mgamc, and against \MadLoop5 in a mode that does not use the polynomial decomposition of the integrand numerator (see appendix~\ref{subsec:newoptions}).
The list of processes cross-checked includes, among others, $d \bar{d}\rightarrow h g g g$ and all gluon fusion processes with up to three identical neutral massive bosons in the final states.
To facilitate future comparisons, we report in appendix~\ref{localRes} the numerical result for a chosen phase-space point of the process $g g \rightarrow h h g g$.
Secondly, at the integrated level, we compared our prediction for the partial decay width $z \rightarrow g g g$ to the result reported in ref.~\cite{vanderBij:1988ac}.\footnote{The result reported in item g.7 of table~\ref{decayTable} differs from the one of ref.~\cite{vanderBij:1988ac} where the value of the top mass, $\alpha_s$ and $\alpha^{-1}$ are chosen to be 173 GeV, 0.134 and 128.0 respectively.} Additional comparisons with the Higgs Effective Theory are presented in sect.~\ref{sec:higgs}. Finally, we tested Lorentz invariance, crossing symmetries and gauge invariance using Ward identities with the standard checks implemented via the command `{\tt check}' of \mgamc\ interface.

\subsection{Phase-space integration and event generation}
\label{sec:PSint}

The phase-space integration and event generation is based on the \madevent~\cite{Maltoni:2002qb} algorithm that we improved in the context of this work. At its core lies the {\emph{diagram enhancement method}} which separates the integration into a sum of integrals whose singularity structure is dictated by a single Feynman diagram topology. Each of these individual integrals, referred to as \emph{integration channel}, is then integrated using an appropriate phase-space parametrisation undoing its underlying structure. For this reason, using \madevent\ phase-space mapping algorithms requires to build tree-level topologies from the contributing loop diagrams. This is achieved by contracting the loop to a single vertex point. It is important to stress here that this mapping to tree topologies is only used to setup the phase-space parametrisation, and at no point to build any sort of numerical estimate of the corresponding loop-induced amplitude (the full exact loop-induced matrix element is used throughout our implementation).

A \madevent\ run involves two steps. The first one is referred to as the \emph{survey} and consists in computing the cross-section for each integration channel down to a given accuracy of 5\%.  Using the information on relative cross-sections and efficiencies provided by the survey, \madevent\ proceeds with a second step referred to as the \emph{refine} where event generation takes place. The time necessary to run the refine step is approximatively linearly proportional to the requested number of unweighted events whilst the one of the survey is independent of this number.

We have implemented a series of improvements to \madevent, the most important of which being the implementation of a dynamical importance sampling for the Monte-Carlo over helicity configurations (\emph{i.e.} the frequency of probing a particular helicity configuration is dynamically adjusted to its relative contribution). Notice however that, as it is the case for standard adaptive Monte-Carlo~\cite{Lepage:1980dq}, we do not account for any correlation between the kinematic variables of integration and the sampling distribution of helicity configurations. The gain obtained thanks to this new operational mode is made explicit in table~\ref{tab:timing}. In that table and for a small set of loop-induced processes, we present the cumulated CPU-hours\footnote{\label{CPUhours}This is equivalent to how long the integration would have taken if it was run sequentially on a single CPU.} necessary for the survey step as well as for the refine to 10k unweighted events. 
We also report the total number of phase-space points for which the matrix-elements needed to be evaluated during the integration.\footnote{Notice that for the processes $p p \rightarrow h j$ and $p p \rightarrow h j j$, the repartition of phase-space points probed across different partonic subprocesses is not the same for the survey and refine steps or between the two different integration techniques. This explains for example the difference of timing per phase-space point between the survey and refine steps.}
We warn the reader that the quantitative results reported here must not be interpreted too literally given that the speed of the node assigned by the cluster can vary from one run to the other.
We compare timings for an integration using a Monte-Carlo sampling with an explicit sum over helicity configurations. In the latter case, we further compare three different reduction methods; OPP as implemented in {\sc\small CutTools} and TIR as implemented in {\sc\small PJFry} or {\sc\small IREGI}.

\begin{table*}[t]
\renewcommand{\arraystretch}{1.3}
\begin{center}
    \begin{tabular}{c|c|c|c}
        \hline \hline
         Helicity sum & Monte-Carlo & \multicolumn{2}{c}{Exact}  \\
        \hline
        Loop Reduction & {\sc\small CutTools} & {\sc\small CutTools}  & TIR \\
        \hline
        \multicolumn{4}{c}{Survey}\\
        \hline
        $ p p \rightarrow h j$ &  13m (125k) &  32m (260k) & {\bf 9m (260k) } \\ 
        $ p p  \rightarrow h j j$ &   {\bf 2d4h (1.2M) }& 16d10h (5.4M) &  9d13h (5.4M)${}^*$  \\
        $ g g \rightarrow z z$  &   {\bf 1h06m (34k) }  & 12h50m (255k) & 1h44m (255k) \\
        $ g g \rightarrow z h g$ &   {\bf 11h13m (110k)}  & 1d8h (516k)  & 1d4h (516k)${}^*$ \\
        \hline
        \multicolumn{4}{c}{Refine}\\
        \hline        
        $ p p \rightarrow h j$ &  1h43m (385k) &  23m (431k)  &  {\bf 6m (431k)} \\
        $ p p  \rightarrow h j j$ &  {\bf 7d17h (2.18M)}  & 75d1h (20.6M)  &  51d19h (20.6M)${}^*$ \\
        $ g g \rightarrow z z$ &  {\bf 7h20m (407k)}  &  4d13h (4.55M)  & 23h07m (5.78M)   \\        
        $ g g \rightarrow z h g$ &  {\bf 23h03m (277k)} & 2d22h (1.13M) & 3d14h (1.4M)${}^*$  \\           
\hline \hline
\end{tabular}
 \caption{\label{tab:timing}  Cumulated CPU-hours necessary for the generation of 10k unweighted events for various loop-induced processes. This corresponds to a Monte-Carlo accuracy on the cross-section typically better than a percent. The numbers in parenthesis specify the number of \emph{polarised} (even when summing exactly over helicity configurations) phase-space points for which the matrix elements needed to be evaluated. The column `TIR' only reports the fastest timing between using the {\sc\small PJFRY} and {\sc\small IREGI} implementation of TIR and we suffixed the timing by a star ($^\star$) when the latter is faster. For each process, we underlined the fastest of all approaches by reporting it in bold font.}
\end{center} 
\end{table*}

We find that when summing exactly over helicity configurations for each phase-space points, TIR is advantageous thanks to the recycling of tensorial coefficients across these configurations (see sect.~\ref{techloop}). However, in the case of processes of larger multiplicity ($pp\rightarrow h j j$) or with many helicity configurations ($g g \rightarrow z z$), this is not sufficient to overcome the gain obtained from sampling over helicity, which considerably reduces the number of polarised phase-space points that needs to be probed to reach a given accuracy.

Given that the typical time for the evaluation of a $2\rightarrow4$ loop-induced matrix element for a phase-space point can already reach several seconds (see table~\ref{tab:MLperf} in appendix~\ref{localRes}), it is of paramount importance to be able to scale \madevent\ parallelisation independently of the number of integration channels. In this way, even though the cumulated sequential CPU-hours necessary for the computation remains constant, the user time spent for the integration of a process can be reduced proportionally to the number of cores available.  Although Monte-Carlo integration methods are intrinsically trivially parallel, adaptive techniques, where the phase-space probing distribution is improved over time, break the mutual independence of two successive iterations.
The original approach of \madevent\ implements the simplest parallelisation in this context, which consists in running each channel of integration as an independent separate job.  
However, this is not sufficient for loop-induced processes which exhibit a small number of time-consuming channels. To push parallelisation further, we have therefore modified the steering of \madevent\  so as to submit multiple jobs on the cluster for each iteration. In between iterations, the phase-space sampling grids and cross-section results from each job are combined together to build the input of the next iteration which is again split into a series of new independent jobs. 

Finally, we have also improved event generation. In \madevent, the unweighting operation follows each iteration and if it did not generate enough unweighted events then a new iteration is submitted --setup to probe twice as many phase-space points-- and the events already generated are simply discarded.
We ameliorated the unweighting step by allowing to combine events generated in any of the previous iterations.
One caveat lies in the fact that if the MC sampling grids improve significantly from one iteration to the other, this procedure does not take advantage of the resulting increase in unweighting efficiency.
We remedy this problem by estimating the remaining computing time for each of the two strategies (discarding or not events from previous iterations) and choosing the most efficient one.

Using a still private implementation of an interface between \MadLoop\ and \Sherpa, we carried a detailed comparison at the integrated level for the loop-induced process $g g \rightarrow Z Z$, including both the triangle (featuring an s-channel Higgs) and box topologies. In particular, we compared to the per-mil accuracy the cross-sections for each individual squared topology as well as for their interference only. We found perfect agreement between \MadEvent\ and \Sherpa. We also checked that integrating all contributions together yields consistent results. Notice that in both computations of this check, the loop-induced matrix element implementation is identical, being the one provided by \MadLoop, and as such this validation targets only \madevent\ integration procedure. Additional validations of the phase-space integration are presented in sect.~\ref{sec:higgs} and~\ref{sec:Validation}. Finally, collaborators have compared our predictions with results for the processes $p p \rightarrow V V$  obtained completely independently~\cite{Kauer:2015dma}.

\newpage
\section{Results in the Standard Model}
\label{sec:results}
In this section, we present results for inclusive cross-sections of loop-induced processes within the Standard Model, whose parameters are set to the values listed in table~\ref{tableParams}.
\begin{table}[ht]
\begin{center}
\begin{tabular}{ll|ll}\midrule\midrule
Parameter & value & Parameter & value
\\\midrule
$\alpha_{S}(m_Z^2)$ & $ \left \{\; \begin{matrix} \text{set by PDF if present}\;(0.13355) \\ 0.118\;\text{otherwise} \end{matrix} \right.  $ & $n_{lf}$ & \tt{4}
\\
$\mu_R=\mu_F$ & $\tt{\hat{\mu}}$ (see text for def.) & $m_{b}=y_{b}$ & \tt{4.7}
\\
$m_{t}=y_{t}$ & \tt{173.0} & $\Gamma_{t}$ & \tt{0}
\\
$G_F$ & \tt{1.16639e-05}  & $\alpha^{-1}$ & \tt{132.507}
\\
$m_Z$ & \tt{91.188} & $\Gamma_{Z}$ & \tt{2.4414}
\\
$m_{W}$ & $\frac{M_Z}{\sqrt{2}}\sqrt{1 + \sqrt{1 - \frac{4\pi}{\sqrt{2}}\frac{\alpha}{G_F M_Z^2}} }$ & $\Gamma_{W}$ & \tt{2.0476}
\\
$m_H$ & \tt{125.0} & $\Gamma_{H}$ & \tt{0.00638}
\\
$V^{CKM}_{ij}$ & $\delta_{ij}$  & $m_{e^\pm}=m_{\mu^\pm}$ & \tt{0.0}
\\
$m_{\tau^\pm}=y_{\tau^\pm}$ & $1.777$  & $\Gamma_{\tau^\pm}$ & \tt{0.0}\\
\midrule\midrule
\end{tabular}
\end{center}
\caption{\label{tableParams} Standard Model parameters used for obtaining the results presented in tables~\ref{ResTab2to1}-\ref{ResTabDecay}. Dimensionful parameters are given in GeV.}
\end{table}

We considered \emph{all} hadronic loop-induced processes in the SM up to three particles in the final states, featuring one, two or three heavy bosons and/or photons as presented in tables~\ref{ResTab2to1},~\ref{ResTab2to2} and~\ref{ResTab2to3} respectively. The seemingly missing processes are all zero, either because of Furry's theorem (such as in $g g \rightarrow h a$ for instance, where all fermionic loops with clockwise and anti-clockwise flow cancel pair-wise) or Landau-Yang's theorem~\cite{Landau:1948kw,Yang:1950rg} that forbids a massive vector particle to decay into two identical massless ones, such as in $g g \rightarrow z$. In addition, we present in table~\ref{ResTab2to4} cross-sections obtained for a selected list of processes with four external final states as well as loop-induced processes with non-hadronic initial states.

For each process, we generated a sample of 10k unweighted events, yielding a Monte-Carlo accuracy on the inclusive cross-section of at least 1\% but typically better. The central factorisation and renormalisation scales are set dynamically to half the sum of all final state transverse energies in the case of scattering processes and statically to the decaying particle mass in the case of decay processes; that is
\begin{equation}
\hat{\mu}= \left \{ \begin{matrix}
m_X, &\:\:\:\: \forall\;\text{ decay proc.}\;X \rightarrow \{x_i\} \\
\frac{1}{2}\sum_{i=n_i}^{n_i+n_f}\sqrt{E^2_{i}+P^2_{T,i}}\;, &\:\:\:\: \text{otherwise}
\end{matrix}\right. .
\end{equation}

Processes with at least one external state Higgs accompanied by only jets and/or photons do not receive any tree-level contributions even in the presence of quarks. For these processes we considered the contributions from the gluon and all four massless quark flavours in the proton and jet definitions, that are denoted by $p$ and $j$. For all other processes, we restricted ourselves to only gluons in the external states since the loop corrections to the corresponding processes with external quarks are formally NNLO. Notice that, unless the loop amplitudes are finite, the virtual-virtual contribution of NNLO corrections can, in principle, not be computed with our implementation because loop integrals are evaluated only up to $\mathcal{O}(\epsilon^0)$ in dimensional regularisation. There is however a proposal~\cite{Weinzierl:2011uz} for a formalism avoiding the computation of the $\mathcal{O}(\epsilon)$ terms of one-loop amplitudes in NNLO computations by capturing these terms with the one-loop insertion operator instead.

For all processes with coloured initial states, we used the MSTW 2008 PDF set~\cite{Martin:2009iq} (with name `{\tt{MSTW2008lo68cl\_nf4}}').
For each process, we indicate the maximum scale variation, denoted by $\Delta_{\hat{\mu}}$, obtained by \emph{independently} multiplying the scales $\mu_r$ and $\mu_f$ by the customary factors one-half, one and two. We also show the PDF uncertainty $\Delta_{PDF}$ obtained from the corresponding MSTW error sets. Both these quantities have been computed from a single run with central PDF set and scale using the exact reweighting approach implemented by the \mgamc\ module \syscalc~\cite{LargeScaleProdPaper}.

We applied the cuts listed in table~\ref{tab:global_cuts} to all processes when applicable.
\begin{table}[ht]
\begin{center}
\begin{small}
\begin{tabular}{c|cl}\midrule\midrule
Cut & Constraint & Comment\\
\hline
$p_{t,j}$   &  $> 20\;$GeV &  $j\;\equiv$ gluons and massless quarks\\
$p_{t,a}, p_{t,l}$  &  $> 10\;$GeV & $a\;\equiv$ photon, $l\;\equiv$ any lepton\\
$\eta_j$   &  $< 5$& $\eta\; \equiv$ rapidity\\
$\eta_a, \eta_l$  &  $< 2.5$& \\
$\Delta R_{j,j}$, $\Delta R_{j,a}$, $\Delta R_{a,a}$, $\Delta R_{l,l}$ &  $> 0.4$& angular separation, $\Delta R = \sqrt{\Delta \phi^2 + \Delta \eta^2}$\\
\midrule\midrule
\end{tabular}                                                                                                                                                                                  
\end{small}
\end{center}
\caption{
\label{tab:global_cuts} General set of cuts applied (when applicable) to all processes of tables~\ref{ResTab2to1}-\ref{ResTabDecay}.
 }
\end{table}

\noindent In addition to these cuts, some processes are subject to more specific changes in the simulation setup:
\begin{itemize}

\item $g g \rightarrow Z Z$, $g g \rightarrow W^+ W^-$ \textbf{(b.6, b.12)} : \\
Cut on heavy boson transverse momenta: $p_{t,V}>1\;$GeV.\\
The fermionic 4-point loop integral with massive identical final state vectors $V$ features an integrable singularity at $p_{t,V}\rightarrow 0$ that we regulated with a technical cut at 1 GeV, alike what is done in the latest version of the code {\sc\small MCFM}~\cite{Campbell:2010ff}.

\item $g g\rightarrow Z W^+W^-$ \textbf{(c.10)}:\\
Complex mass scheme with $\Gamma_{top}=1.49\;$GeV.\\
Some of the diagrams contributing to this process can have up to four loop propagators onshell\footnote{An example of which is the pentagon with three top quark loop propagators and one bottom quark one, from which the Z-boson is emitted.}, and some of the resulting four-point scalar loop integrals are unstable for kinematic configurations with a total invariant mass slightly above twice the top mass.

To avoid this issue, we considered here a non-zero top quark width within the complex mass scheme~\cite{Denner:2006ic,Denner:2005fg}, but keeping the weak boson widths set to zero since they are onshell in the external states. Such a partial assignment of the widths within the complex mass scheme is consistent at leading-order when there is no intermediate resonant weak bosons. In this context, the only difference w.r.t. the narrow width approximation is that the top quark Yukawa coupling as well as the mass in the numerator of top quark propagators are complex.
 
\item $e^+ e^-\rightarrow H H$ \textbf{(e.2)}:\\
The widths of the massive bosons are set to zero in the loop propagators but kept finite (and equal to the values given in table~\ref{tableParams}) in the propagators of the tree structures attached to the loops.\\
Contrary to the two other leptonic scattering processes presented in this section, this one does not involve fermionic loops but genuine weak loops. We stress that the complex mass scheme would be the optimal approach for accounting for finite-widths effect since it yields gauge-invariant results.\footnote{This application of the complex mass scheme within EW loop computations will be made possible soon within the \mgamc\ framework~\cite{CMSMGPAPER}. Notice however that, being UV-finite, the case of loop-induced processes is considerably simpler for what concerns the application of the complex mass scheme.}\\
 
\item $p p\rightarrow t t$ \textbf{(f.1)}:\\
\begin{figure}
	\centering
	\includegraphics[clip,scale=0.8]{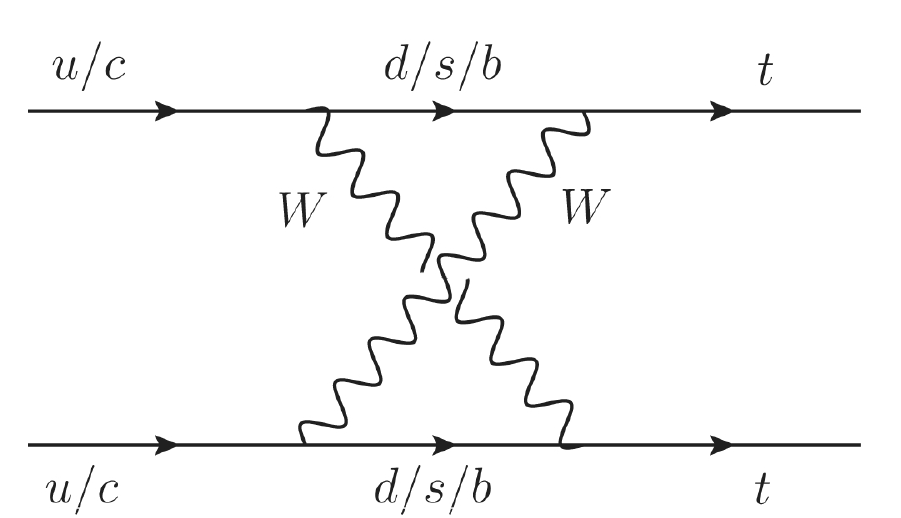}
	\caption{\label{fig:ttdiags} Representative diagram for the pair-production of same-sign top quarks in the SM. }
\end{figure}
This process is extremely rare in the SM for three reasons: it only occurs via a box loop with two $W$-boson propagators (see Fig.~\ref{fig:ttdiags}) and it is suppressed by the square of CKM matrix off-diagonal elements as well as the bottom quark mass because of the GIM mechanism. Observing such final states would therefore be a clear indication of new physics. We chose the following values for the additional parameters entering this computation:\\
$\Gamma_{W}=0$ and the CKM matrix set to non-unity, expressed in terms of the Wolfenstein parametrization~\cite{Wolfenstein:1983yz}, with $\lambda=0.2253$, $A=0.808$, $\rho=0.132$ and $\eta=0.341$.

\item Decay processes (\textbf{g.x}):\\
Processes \textbf{g.1}, \textbf{g.4} and \textbf{g.7} are computed completely inclusively.\\
For processes \textbf{g.2}, \textbf{g.3}, \textbf{g.5} and \textbf{g.7}, all pairs of identical final state objects are restricted to have an invariant mass $\Delta m_{jj/aa}>10$ GeV and all decay products must have an angular separation of $\Delta R_{j/a,j/a}>0.4$. 

\end{itemize}

In tables~\ref{ResTab2to1}-\ref{ResTabDecay}, we denote by a dagger ($^\dagger$) processes whose inclusive cross-section has never been published and is, to the best of our knowledge, reported here for the first time. We prefix by a star ($^\star$) processes which are not readily available in the main public simulation tools {\sc \small MCFM}~\cite{Campbell:2010ff}, {\sc \small VBFNLO}~\cite{Arnold:2008rz} or {\sc \small HPAIR}~\cite{Plehn:1996wb}.

\begin{table}[h]
\begin{center}
\begin{small}
\begin{tabular}{l r@{$\,\to\,$}l lll c}
\midrule\midrule
\multicolumn{3}{c}{Process~~~~~~~~~~~~~~~~~~~}& Syntax  & \multicolumn{1}{c}{\hspace{-0.25cm}Cross section (pb) \hspace{0.9cm} $\Delta_{\hat{\mu}}$\hspace{2mm} $\Delta_{PDF}$} & Ref. \\
\multicolumn{3}{c}{Single boson + jets~~~}& &  \multicolumn{1}{c}{$\sqrt s=$ 13 TeV} \\
\midrule
a.1 & $pp$ & $H$  & {\tt p p > h [QCD] } &$ 17.79 \pm 0.060\phantom{\, \cdot 10^{+0}d}  \quad {}^{+31.3 \% }_{-23.1 \% } \,\, {}^{+0.5 \% }_{-0.9 \% }$ & \cite{Heinemeyer:2013tqa} \\a.2 & $pp$ & $Hj$    & {\tt p p > h j [QCD]}  &$ 12.86 \pm 0.030\phantom{\, \cdot 10^{+0}d}  \quad {}^{+42.3 \% }_{-27.7 \% } \,\, {}^{+0.6 \% }_{-0.9 \% }$ & \cite{Heinemeyer:2013tqa}\\a.3 & $pp$ & $Hjj$   & {\tt p p > h j j QED=1 [QCD]}  &$ 6.175 \pm 0.020\phantom{\, \cdot 10^{+0}d}  \quad {}^{+61.8 \% }_{-35.6 \% } \,\, {}^{+0.7 \% }_{-0.9 \% }$ & \cite{Heinemeyer:2013tqa}\\
\midrule
${}^\star$a.4 & $gg$ & $Zg$   & {\tt g g > z g  [QCD]} &$ 43.05 \pm 0.060\phantom{\, \cdot 10^{+0}d}  \quad {}^{+43.7 \% }_{-28.4 \% } \,\, {}^{+0.7 \% }_{-1.0 \% }$ & \cite{vanderBij:1988ac}\\${}^\star$a.5 & $gg$ & $Zgg$  & {\tt g g > z g g [QCD]} &$ 20.85 \pm 0.030\phantom{\, \cdot 10^{+0}d}  \quad {}^{+64.5 \% }_{-36.5 \% } \,\, {}^{+1.0 \% }_{-1.1 \% }$  &\cite{Kidonakis:2014zva}\\
\midrule
${}^\dagger$a.6 & $gg$ & $\gamma g$ & {\tt g g > a g [QCD]}  &$ 75.61 \pm 0.200\phantom{\, \cdot 10^{+0}d}  \quad {}^{+73.8 \% }_{-41.6 \% } \,\, {}^{+0.7 \% }_{-1.1 \% }$ & [\;--\;]\\${}^\dagger$a.7 & $gg$ & $\gamma gg$ & {\tt g g > a g g [QCD]} &$ 14.50 \pm 0.030\phantom{\, \cdot 10^{+0}d}  \quad {}^{+76.2 \% }_{-40.7 \% } \,\, {}^{+0.6 \% }_{-1.0 \% }$ & [\;--\;]\\
\midrule\midrule
\end{tabular}
\end{small}
\end{center}
\caption{
\label{ResTab2to1}
Inclusive cross-sections for loop-induced single electroweak boson production in association with up to two jets/gluons. A star ($^\star$) prefixes processes not readily available in the tools {\sc \small MCFM}, {\sc \small VBFNLO} or {\sc \small HPAIR}. A dagger ($^\dagger$) prefixes processes whose inclusive cross-section is reported here for the first time. See text for details.}
\end{table}

\begin{table}[h]
\begin{center}
\begin{small}
\begin{tabular}{l r@{$\,\to\,$}l lll c }
\midrule\midrule
\multicolumn{3}{c}{Process~~~~~~~~~~~~~~~~~~~}& Syntax  & \multicolumn{1}{c}{\hspace{-0.25cm}Cross section (pb) \hspace{0.9cm} $\Delta_{\hat{\mu}}$\hspace{2mm} $\Delta_{PDF}$} & Ref.\\
\multicolumn{3}{c}{Double bosons + jet~~}& &  \multicolumn{1}{c}{$\sqrt s=$ 13 TeV} \\
\midrule
b.1 & $pp$ & $HH$  & {\tt p p > h h [QCD] } &$ 1.641 \pm 0.002\, \cdot 10^{-2}  \quad {}^{+30.2 \% }_{-21.7 \% } \,\, {}^{+1.1 \% }_{-1.2 \% }$ & \cite{Plehn:1996wb}\\b.2 & $pp$ & $HHj$    & {\tt p p > h h j [QCD]}  &$ 1.758 \pm 0.003\, \cdot 10^{-2}  \quad {}^{+45.7 \% }_{-29.2 \% } \,\, {}^{+1.2 \% }_{-1.2 \% }$ & \cite{Dolan:2012rv}\\${}^\star$b.3 & $pp$ & $H\gamma j$    & {\tt p p > h a j [QCD]}  &$ 4.225 \pm 0.006\, \cdot 10^{-3}  \quad {}^{+38.6 \% }_{-25.9 \% } \,\, {}^{+0.4 \% }_{-0.7 \% }$ & \cite{Agrawal:2014tqa}\\${}^\star$b.4 & $gg$ & $HZ$  & {\tt g g > h z [QCD] } &$ 6.537 \pm 0.030\, \cdot 10^{-2}  \quad {}^{+29.4 \% }_{-21.3 \% } \,\, {}^{+1.0 \% }_{-1.1 \% }$ & \cite{Ferrera:2014lca}\\${}^\star$b.5 & $gg$ & $HZg$    & {\tt g g > h z g [QCD]}  &$ 5.465 \pm 0.020\, \cdot 10^{-2}  \quad {}^{+46.0 \% }_{-29.4 \% } \,\, {}^{+1.2 \% }_{-1.3 \% }$& \cite{Agrawal:2014tqa}\\
\midrule
b.6 & $gg$ & $ZZ$  & {\tt g g > z z [QCD] } &$ 1.313 \pm 0.004\phantom{\, \cdot 10^{+0}d}  \quad {}^{+27.1 \% }_{-20.1 \% } \,\, {}^{+0.7 \% }_{-1.0 \% }$ & \cite{Campbell:2010ff}\\${}^\star$b.7 & $gg$ & $ZZg$    & {\tt g g > z z g [QCD]}  &$ 0.6361 \pm 0.002\phantom{ 10^{+0}d}  \quad {}^{+45.4 \% }_{-29.1 \% } \,\, {}^{+1.0 \% }_{-1.2 \% }$ & \cite{Campanario:2012bh}\\b.8 & $gg$ & $Z\gamma$  & {\tt g g > z a [QCD] } &$ 1.265 \pm 0.0007\phantom{ 10^{+0}d}  \quad {}^{+30.2 \% }_{-22.2 \% } \,\, {}^{+0.6 \% }_{-1.0 \% }$ & \cite{Campbell:2010ff} \\${}^\star$b.9 & $gg$ & $Z\gamma g$    & {\tt g g > z a g [QCD]}  &$ 0.4604 \pm 0.001\phantom{ 10^{+0}d}  \quad {}^{+43.7 \% }_{-28.4 \% } \,\, {}^{+0.8 \% }_{-1.1 \% }$ & \cite{Grazzini:2015nwa}\\
\midrule
b.10 & $gg$ & $\gamma \gamma$  & {\tt g g > a a [QCD]}  &$ 5.182 \pm 0.010\, \cdot 10^{+2}  \quad {}^{+72.3 \% }_{-43.4 \% } \,\, {}^{+1.0 \% }_{-1.3 \% }$ & \cite{Campbell:2010ff}\\${}^\star$b.11 & $gg$ & $\gamma \gamma g$ & {\tt g g > a a g [QCD]}  &$ 19.22 \pm 0.030\phantom{\, \cdot 10^{+0}d}  \quad {}^{+59.7 \% }_{-35.7 \% } \,\, {}^{+0.7 \% }_{-1.0 \% }$ & \cite{Martin:2013ula}\\\midrule
b.12 & $gg$ & $W^+ W^- $  & {\tt g g > w+ w- [QCD]}  &$ 4.099 \pm 0.010\phantom{\, \cdot 10^{+0}d}  \quad {}^{+26.5 \% }_{-19.7 \% } \,\, {}^{+0.7 \% }_{-1.0 \% }$ & \cite{Campbell:2011cu}\\${}^\star$b.13 & $gg$ & $W^+ W^- g$ & {\tt g g > w+ w- g [QCD]}  &$ 1.837 \pm 0.004\phantom{\, \cdot 10^{+0}d}  \quad {}^{+45.2 \% }_{-29.0 \% } \,\, {}^{+0.9 \% }_{-1.1 \% }$ & \cite{Cascioli:2013gfa}\\\midrule\midrule
\end{tabular}
\end{small}
\end{center}
\caption{
\label{ResTab2to2}
Inclusive cross-sections for loop-induced double electroweak boson production in association with up to one jet/gluon. A star ($^\star$) prefixes processes not readily available in the tools {\sc \small MCFM}, {\sc \small VBFNLO} or {\sc \small HPAIR}. A dagger ($^\dagger$) prefixes processes whose inclusive cross-section is reported here for the first time. See text for details.}
\end{table}


\begin{table}[ph]
\begin{center}
\begin{small}
\begin{tabular}{l r@{$\,\to\,$}l lll c}
\midrule\midrule
\multicolumn{3}{c}{Process~~~~~~~~~~~~~~~~~~~}& Syntax  & \multicolumn{1}{c}{\hspace{-0.25cm}Cross section (pb) \hspace{0.9cm} $\Delta_{\hat{\mu}}$\hspace{2mm} $\Delta_{PDF}$} & Ref.\\
\multicolumn{3}{c}{Triple bosons~~~~~~~~~~~~}& &  \multicolumn{1}{c}{$\sqrt s=$ 13 TeV} \\
\midrule
${}^\star$c.1 & $pp$ & $HHH$  & {\tt p p > h h h [QCD] } &$ 3.968 \pm 0.010\, \cdot 10^{-5}  \quad {}^{+31.8 \% }_{-22.6 \% } \,\, {}^{+1.4 \% }_{-1.4 \% }$ & \cite{Maltoni:2014eza}\\${}^\dagger$c.2 & $gg$ & $HHZ$    & {\tt g g > h h z [QCD]}  &$ 5.260 \pm 0.009\, \cdot 10^{-5}  \quad {}^{+31.2 \% }_{-22.2 \% } \,\, {}^{+1.3 \% }_{-1.3 \% }$ & [\;--\;]\\${}^\dagger$c.3 & $gg$ & $HZZ$  & {\tt g g > h z z [QCD] } &$ 1.144 \pm 0.004\, \cdot 10^{-4}  \quad {}^{+31.1 \% }_{-22.2 \% } \,\, {}^{+1.2 \% }_{-1.3 \% }$ & [\;--\;]\\${}^\dagger$c.4 & $gg$ & $HZ\gamma$    & {\tt g g > h z a [QCD]}  &$ 6.190 \pm 0.020\, \cdot 10^{-6}  \quad {}^{+29.3 \% }_{-21.2 \% } \,\, {}^{+1.0 \% }_{-1.2 \% }$ & [\;--\;]\\${}^\dagger$c.5 & $pp$ & $H\gamma\gamma$    & {\tt p p > h a a [QCD]}  &$ 6.058 \pm 0.004\, \cdot 10^{-6}  \quad {}^{+30.3 \% }_{-21.8 \% } \,\, {}^{+1.1 \% }_{-1.3 \% }$ & [\;--\;]\\${}^\star$c.6 & $gg$ & $HW^+W^-$    & {\tt g g > h w+ w- [QCD]}  &$ 2.670 \pm 0.007\, \cdot 10^{-4}  \quad {}^{+31.0 \% }_{-22.2 \% } \,\, {}^{+1.2 \% }_{-1.3 \% }$  & \cite{Mao:2009jp}\\
\midrule
${}^\dagger$c.7 & $gg$ & $ZZZ$    & {\tt g g > z z z [QCD]}  &$ 6.964 \pm 0.009\, \cdot 10^{-5}  \quad {}^{+30.9 \% }_{-22.1 \% } \,\, {}^{+1.2 \% }_{-1.3 \% }$  & [\;--\;]\\${}^\dagger$c.8 & $gg$ & $ZZ\gamma$  & {\tt g g > z z a [QCD] } &$ 3.454 \pm 0.010\, \cdot 10^{-6}  \quad {}^{+28.7 \% }_{-20.9 \% } \,\, {}^{+0.9 \% }_{-1.1 \% }$ & [\;--\;]\\${}^\star$c.9 & $gg$ & $Z\gamma\gamma$    & {\tt g g > z a a [QCD]}  &$ 3.079 \pm 0.005\, \cdot 10^{-4}  \quad {}^{+28.0 \% }_{-20.9 \% } \,\, {}^{+0.7 \% }_{-1.0 \% }$ & \cite{Agrawal:2012as}\\${}^\dagger$c.10 & $gg$ & $ZW^+W^-$    & {\tt g g > z w+ w- [QCD]}  &$ 8.595 \pm 0.020\, \cdot 10^{-3}  \quad {}^{+26.9 \% }_{-19.5 \% } \,\, {}^{+0.6 \% }_{-0.6 \% }$ & [\;--\;]\\
\midrule
${}^\dagger$c.12 & $gg$ & $\gamma W^+ W^-$  & {\tt g g > a w+ w- [QCD] } &$ 1.822 \pm 0.005\, \cdot 10^{-2}  \quad {}^{+28.7 \% }_{-20.9 \% } \,\, {}^{+0.9 \% }_{-1.1 \% }$ & [\;--\;]\\
\midrule\midrule
\end{tabular}
\end{small}
\end{center}
\caption{
\label{ResTab2to3}
Inclusive cross-sections for loop-induced triple electroweak boson production. A star ($^\star$) prefixes processes not readily available in the tools {\sc \small MCFM}, {\sc \small VBFNLO} or {\sc \small HPAIR}. A dagger ($^\dagger$) prefixes processes whose inclusive cross-section is reported here for the first time. See text for details.}
\end{table}

\begin{table}[ph]
\label{ResTab2to4}	
\begin{center}
\begin{small}
\begin{tabular}{llllll c }
\midrule\midrule
\multicolumn{2}{c}{Process~~~~~~~~~~~~~~~~~~~}& Syntax  & \multicolumn{1}{c}{\hspace{-0.25cm}Cross section (pb) \hspace{0.9cm} $\Delta_{\hat{\mu}}$\hspace{2mm} $\Delta_{PDF}$} & Ref.\\
\multicolumn{2}{c}{Selected $2\rightarrow4$~~~~~~~~~}& &  \multicolumn{1}{c}{$\sqrt s=$ 13 TeV} \\
\midrule
${}^\dagger$d.1 & $p p\rightarrow Hjjj$  & {\tt \scriptsize{p p > h j j j QED=1 [QCD] }} &$ 2.519 \pm 0.005\phantom{\, \cdot 10^{+0}d} \quad {}^{+75.1 \% }_{-39.8 \% } \,\, {}^{+0.6 \% }_{-0.6 \% }$ & \cite{Campanario:2013mga}\\${}^\star$d.2 & $p p\rightarrow HHjj$  & {\tt \scriptsize{p p > h h j j QED=1 [QCD] }} &$ 1.085 \pm 0.002\, \cdot 10^{-2}  \quad {}^{+62.1 \% }_{-35.8 \% } \,\, {}^{+1.2 \% }_{-1.3 \% }$ & \cite{Dolan:2013rja} \\${}^\dagger$d.3 & $p p\rightarrow HHHj$  & {\tt \scriptsize{p p > h h h j [QCD] }} &$ 4.981 \pm 0.008\, \cdot 10^{-5}  \quad {}^{+46.3 \% }_{-29.6 \% } \,\, {}^{+1.4 \% }_{-1.4 \% }$& [\;--\;]\\${}^\dagger$d.3 & $p p\rightarrow HHHH$  & {\tt \scriptsize{p p > h h h h [QCD] }} &$ 1.080 \pm 0.003\, \cdot 10^{-7}  \quad {}^{+33.3 \% }_{-23.4 \% } \,\, {}^{+1.7 \% }_{-1.7 \% }$& [\;--\;]\\d.4 & \scriptsize{$g g\rightarrow e^+e^-\mu^+\mu^-$}    & {\tt \scriptsize{g g > e+ e- mu+ mu- [QCD]}}  &$ 2.022 \pm 0.003\, \cdot 10^{-3}  \quad {}^{+26.4 \% }_{-19.4 \% } \,\, {}^{+0.7 \% }_{-1.1 \% }$ & \cite{Kauer:2012hd} \\${}^\dagger$d.5 & $p p\rightarrow HZ\gamma j$    & {\tt \scriptsize{g g > h z a g [QCD]}}  &$ 4.950 \pm 0.008\, \cdot 10^{-6}  \quad {}^{+45.8 \% }_{-29.3 \% } \,\, {}^{+1.2 \% }_{-1.3 \% }$& [\;--\;]\\
\midrule
\multicolumn{2}{c}{Non-hadronic processes}& &  \multicolumn{1}{c}{$\sqrt s=$ 500 GeV, no PDF} \\
\midrule
${}^\star$e.1 & $e^+ e^-\rightarrow ggg$    & \scriptsize{\tt e+ e- > g g g [QED]}  &$ 2.526 \pm 0.004\, \cdot 10^{-6}\phantom{^1}  \quad {}^{+31.2 \% }_{-22.0 \% } \,\, {}^{\phantom{0.0\%} }_{\phantom{0.0\%} }$ & \cite{Weinzierl:2008iv}\\${}^\dagger$e.2 & $e^+ e^-\rightarrow HH$  & \scriptsize{\tt e+ e- > h h [QED] } &$ 1.567 \pm 0.003\, \cdot 10^{-5}\phantom{^1}  \phantom{\quad {}^{+0.0 \% }_{-0.0 \% }} \,\, {}^{\phantom{0.0\%} }_{\phantom{0.0\%} }$& [\;--\;]\\${}^\dagger$e.3 & $e^+ e^-\rightarrow HHgg$  & \scriptsize{\tt e+ e- > h h g g [QED] } &$ 6.629 \pm 0.010\, \cdot 10^{-11}  \quad {}^{+19.2 \% }_{-14.8 \% } \,\, {}^{\phantom{0.0\%} }_{\phantom{0.0\%} }$& [\;--\;]\\${}^\star$e.4 & $\gamma \gamma\rightarrow HH$  & \scriptsize{\tt a a > h h [QED] } &$ 3.198 \pm 0.005\, \cdot 10^{-4}\phantom{^1} \phantom{\quad {}^{+0.0 \% }_{-0.0 \% }} \,\, {}^{\phantom{0.0\%} }_{\phantom{0.0\%} }$ & \cite{Jikia:1992mt}\\
\midrule
\multicolumn{2}{c}{Miscellaneous~~~~~~~~~~~~~}& &  \multicolumn{1}{c}{$\sqrt s=$ 13 TeV} \\
\midrule
${}^\dagger$f.1 & $p p\rightarrow t t$    & {\tt p p > t t [QED]}  &$ 4.045 \pm 0.007\, \cdot 10^{-15}  \quad {}^{+0.2 \% }_{-0.8 \% } \,\, {}^{+0.9 \% }_{-1.0 \% }$& [\;--\;]\\
\midrule\midrule
\end{tabular}
\end{small}
\end{center}
\caption{
\label{ResTab2to4}
Inclusive cross-sections for various $2\rightarrow 4$ processes as well as processes with non-hadronic initial states. A star ($^\star$) prefixes processes not readily available in the tools {\sc \small MCFM}, {\sc \small VBFNLO} or {\sc \small HPAIR}. A dagger ($^\dagger$) prefixes processes whose inclusive cross-section is reported here for the first time. See text for details.}
\end{table}

\begin{table}[ht]
\begin{center}
\begin{small}
\begin{tabular}{l r@{$\,\to\,$}l lll c }
\midrule\midrule
\multicolumn{3}{c}{Process~~~~~~~~~~~~~~~~~~~}& Syntax  & \multicolumn{1}{c}{\hspace{-0.25cm}Partial width (GeV)\hspace{2mm}} & Ref. \\
\multicolumn{3}{c}{Bosonic decays~~~~~~~~~~~~~~~~}& & \\
\midrule
g.1 & $H$ & $j j$  & {\tt h > j j [QCD] } &$ 1.740 \pm 0.0006\, \cdot 10^{-4}$ & \cite{Heinemeyer:2013tqa}\\${}^\star$g.2 & $H$ & $j j j$    & {\tt h > j j j [QCD]}  &$ 3.413 \pm 0.010\, \cdot 10^{-4}$ & \cite{Heinemeyer:2013tqa} \\${}^\dagger$g.3 & $H$ & $j j j j$   & {\tt h > j j j j QED=1 [QCD]}  &$ 1.654 \pm 0.004\, \cdot 10^{-4}$& [\;--\;]\\
\midrule
g.4 & $H$ & $\gamma \gamma$    & {\tt h > a a [QED]}  &$ 9.882 \pm 0.002\, \cdot 10^{-6}$ & \cite{Marciano:2011gm}\\${}^\dagger$g.5 & $H$ & $\gamma \gamma j j$    & {\tt h > a a j j [QCD]}  &$ 7.448 \pm 0.030\, \cdot 10^{-13}$& [\;--\;]\\
${}^\dagger$g.7 & $H$ & $\gamma \gamma \gamma \gamma$    & {\tt h > a a a a [QED]}  &$ 1.546 \pm 0.006\, \cdot 10^{-14}$ & [\;--\;]\\
\midrule
${}^\star$g.8 & $Z$ & $g g g$    & {\tt z > g g g [QCD]}  &$ 3.986 \pm 0.010\, \cdot 10^{-6}$ & \cite{vanderBij:1988ac}\\\midrule\midrule
\end{tabular}
\end{small}
\end{center}
\caption{
\label{ResTabDecay}
Partial decay widths for selected decay processes occurring via loops only. A star ($^\star$) prefixes processes not readily available in the tools {\sc \small MCFM}, {\sc \small VBFNLO} or {\sc \small HPAIR}. A dagger ($^\dagger$) prefixes processes whose inclusive cross-section is reported here for the first time. See text for details.}
\label{decayTable}
\end{table}

\section{A close-up on Higgs production}
\label{sec:higgs}

The purpose of this section is to illustrate the capabilities of \mgamc\ in the context of loop-induced processes and to present additional validation material. All results in this section are for the LHC 13 TeV using generation level cuts and SM parameters identical to those of sect.~\ref{sec:results}.
We focus here on Higgs production via gluon fusion and compare it to the Effective Field Theory (EFT) approach.
This theory approximates the loop by the effective operator:
\begin{equation}
\label{eq:effectivelagrangian}
{\cal L}_{\textrm{eff}} = -\frac{C}{4} H \, G_{\mu \nu }^a G^{a \mu \nu}\,,
\end{equation}
where $H$ is the Higgs field, $G_{\mu \nu }^a$ is the gluon field strength tensor and $C$ is the Wilson coefficient (known up to N${}^4$LO)~\cite{Chetyrkin:1997un, Schroder:2005hy, Chetyrkin:2005ia}. This approximation was recently used in the context of the computation of the Higgs cross-section at N${}^3$LO QCD accuracy, exhibiting a scale uncertainty of 2\% only~\cite{Anastasiou:2015vya}. 
At this level of  precision it is important to consider the corrections to the EFT approach, by including effects due to the finite top quark mass~\cite{Harlander:2009mq,Pak:2009dg,DelDuca:2001eu} as well as the interference between top and bottom quark loops~\cite{Grazzini:2013mca,Dittmaier:2012vm}. 

\begin{figure}
	\centering
	\includegraphics[trim=10 10 40 20,clip,scale=0.40]{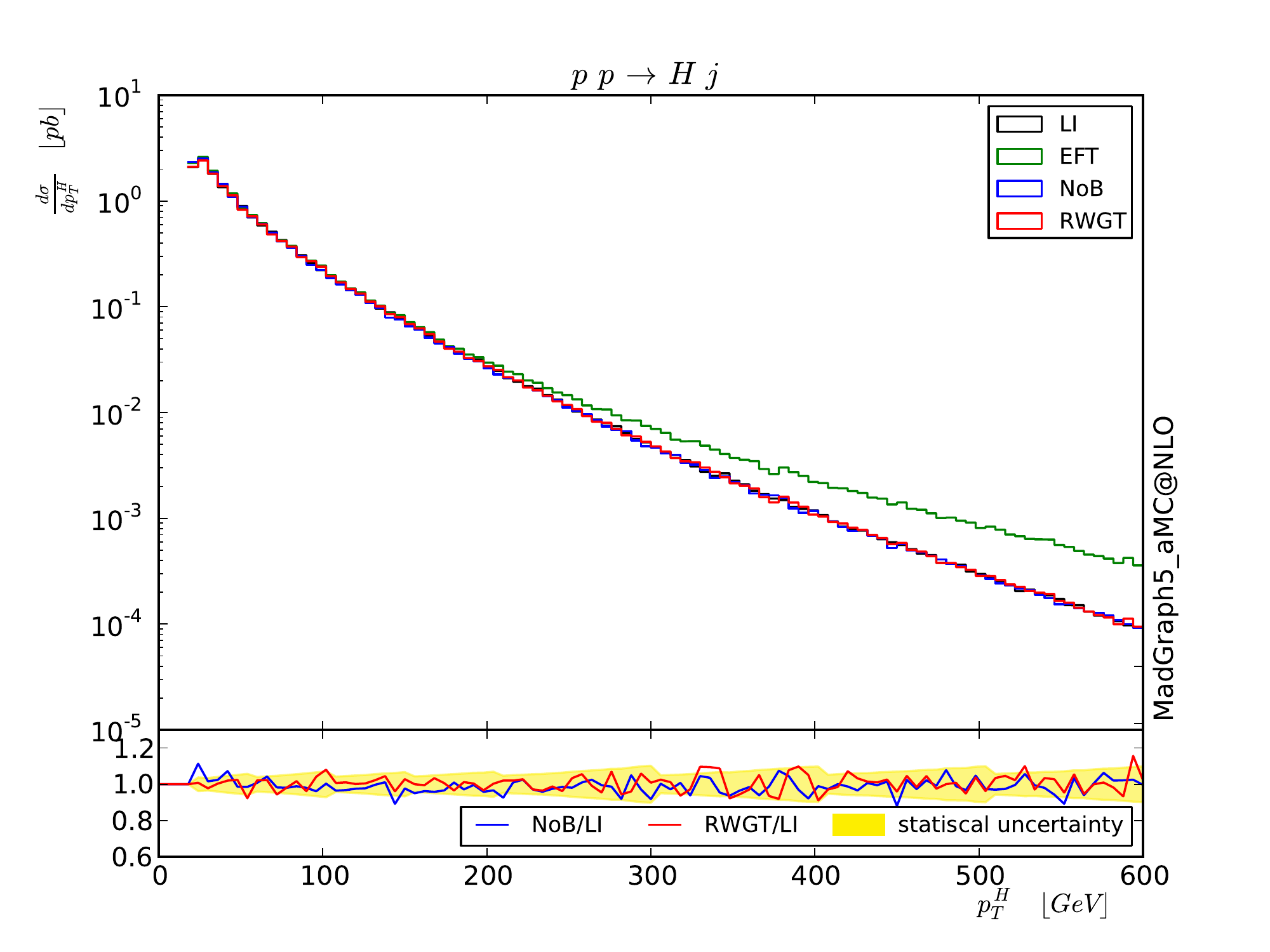}
	\includegraphics[trim=10 10 40 20,clip,scale=0.40]{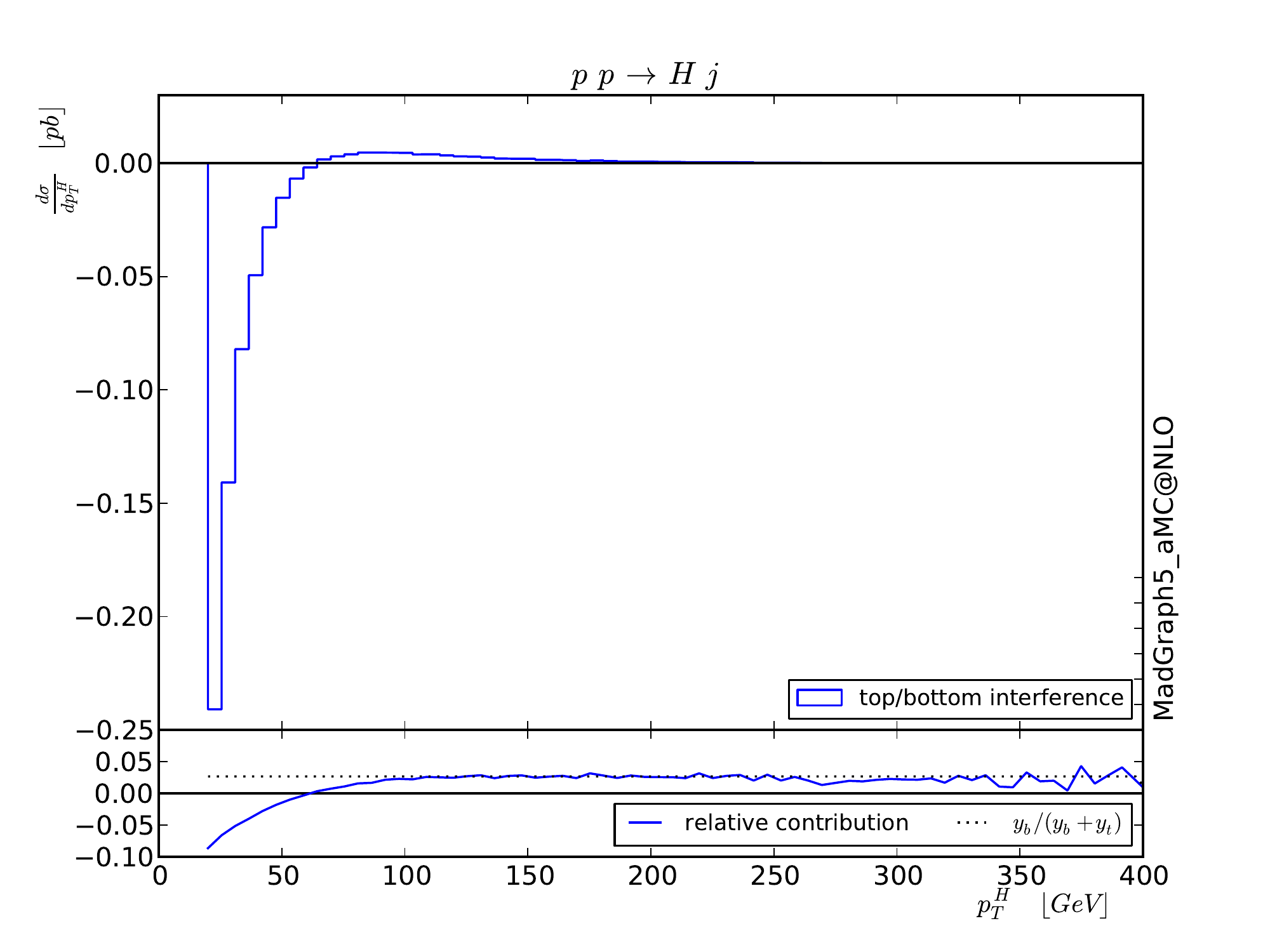}
	\caption{\label{fig:Higgs_pt} Differential distribution of the Higgs transverse momentum. The left panel presents the prediction obtained within the three different computational setups LI, EFT and NoB introduced in table~\ref{tab:cross_ggh}. The red curve, labelled `RWGT', is obtained by reweighting events generated from the effective theory approach with the exact loop-induced matrix element including all top- and bottom quark effects (therefore strictly equivalent to the `LI' prediction in the infinite statistics limit). The bottom inset of the left panel shows the ratios between the various predictions, overlaid (in yellow) by the one sigma statistical error band. The right panel shows the contribution of the interference between the top and bottom quark loops, and its bottom inset indicates its contribution relative to the total cross-section.}
\end{figure}

The importance of these contributions is already significant at leading order accuracy.
In table~\ref{tab:cross_ggh}, we compare the LO cross-section for the Effective Field Theory approach (EFT)\footnote{We use the built-in `{\tt heft}' model of \mgamc~\cite{Alwall:2007st}.} with the SM exact loop-induced process in two scenarios: first with massive bottom quark (labelled `LI') and with massless bottom quark (labelled `NoB') where the contribution of the bottom quark loop is vanishing (both when interfering with the top quark loop and when squared against itself).
For the zero jet multiplicity, we have a perfect agreement between the `NoB' case and the `EFT' one because the effective field theory is valid when $\sqrt{\hat s}=m_H<2m_t$.\footnote{This very good agreement is also due to the presence of mass correction factors in the `{\tt heft}' model.} In presence of radiation, the total energy of the event can lie outside the validity range of the EFT theory and this explains the difference observed between these two schemes. Fig.~\ref{fig:Higgs_pt} (left part) shows the Higgs $p_T$ distribution for various computational setups and it clearly demonstrates that the EFT breaks down for large Higgs transverse momenta.

\begin{table}
\begin{center}                                                                                                                                                                                     
\begin{small}
\begin{tabular}{c|c|c|c}\midrule\midrule
process & EFT  & Exact loop-Induced (LI)  & Exact loop-Induced $m_b=0$ (NoB) \\
\hline
$ g g \to h $                  & 19.996(4) pb  & 17.79(6) pb & 19.94(4) pb\\
$ p p  \to h j $               &  13.41(2)  pb& 12.86(4) pb&  13.24(4) pb\\
$ p p  \to h j j $ & 6.31(2) pb  & 6.18(2) pb &   6.13(1) pb\\
\midrule\midrule
\end{tabular}                                                                                                                                                                                         
\end{small}
\end{center}
\caption{
\label{tab:cross_ggh} Comparison of the cross-section for loop-induced Higgs production computed in different setups. The first column reports the prediction from the Effective Field Theory (EFT) limit using the `{\tt heft}' model~\cite{Alwall:2007st}. The cross-sections of the second column (LI) are computed by directly integrating the loop-induced diagrams, hence keeping all mass effects. The setup of the last column (NoB) is identical to that of the second one, except for the fact that the contribution from bottom quark loops is removed. We do not include the vector boson fusion contribution in the cross-sections reported for the process $p p \rightarrow H j j$.
 }
\end{table}

Another effect emphasised in table~\ref{tab:cross_ggh} is the importance of the interference of the top and bottom quark loops. For the zero jet multiplicity case, this interference is negative because of the absorptive part of the bottom quark loop which is below threshold, unlike the purely real top quark loop. In presence of additional jets, the interference can be either positive or negative, leading to a milder effect on the cross-section but affecting the shape of the distribution. In Fig.~\ref{fig:Higgs_pt} (right part), we show the interference contribution as a function of the Higgs transverse momentum (for the process $p p \rightarrow H j$). 
At low $p_T$, we recover the behaviour of the 0-jet multiplicity case featuring a negative interference effect which contributes to 10\% of the total cross-section. Conversely, at high $p_T$, the interference of the bottom and top quark loop diagrams is constructive and its relative contribution tends to a constant equal to the ratio of the corresponding Yukawa couplings: $\frac{y_b}{y_b+y_t}=2.7\%$.

The left panel of fig.~\ref{fig:Higgs_pt} features a fourth prediction labelled `RWGT', short for \emph{reweighting}. This prediction corresponds to the differential cross-section predicted by the Effective Field theory where the weight of each event has been rescaled by $ \frac{|\mathcal{M}^h_{LI}|^2}{|\mathcal{M}^h_{EFT}|^2}$, with $\mathcal{M}^h_{LI}$ being the loop-induced matrix element for the helicity configuration $h$ computed for the kinematic configuration of the event considered. With infinite statistics, the prediction based on reweighting must be identical to the one obtained from a direct integration of the loop-induced matrix element. We check that the ratio between these two predictions (shown in the bottom inset) is always compatible with one, given the statistical uncertainty whose one-sigma variation is given as the yellow band. This represents a non trivial validation of \MadEvent\ phase-space integration for loop-induced processes.

\begin{figure}
	\centering
	\includegraphics[trim=15 0 30 5,clip,scale=0.4]{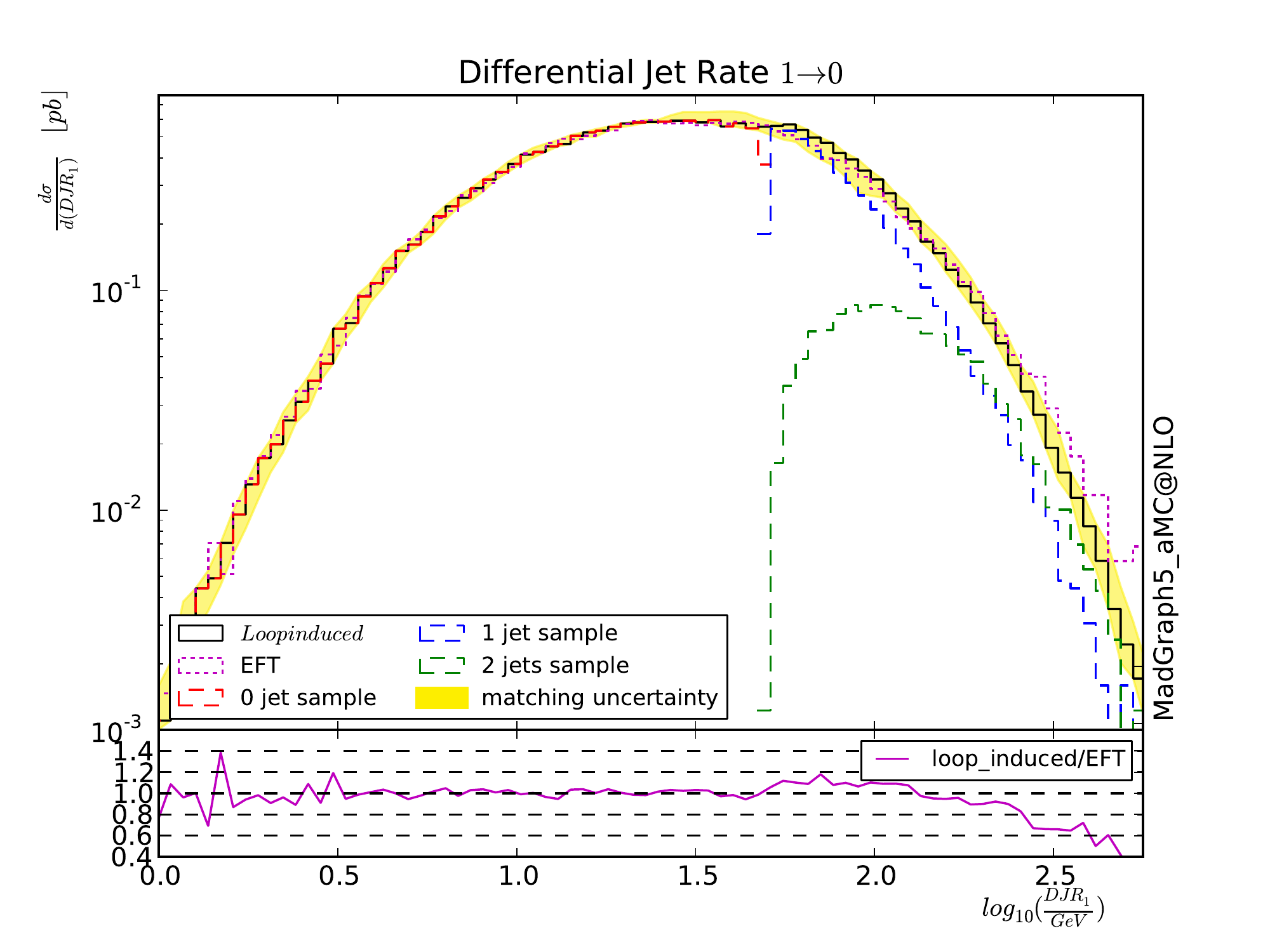}
	\includegraphics[trim=15 0 30 5,clip,scale=0.4]{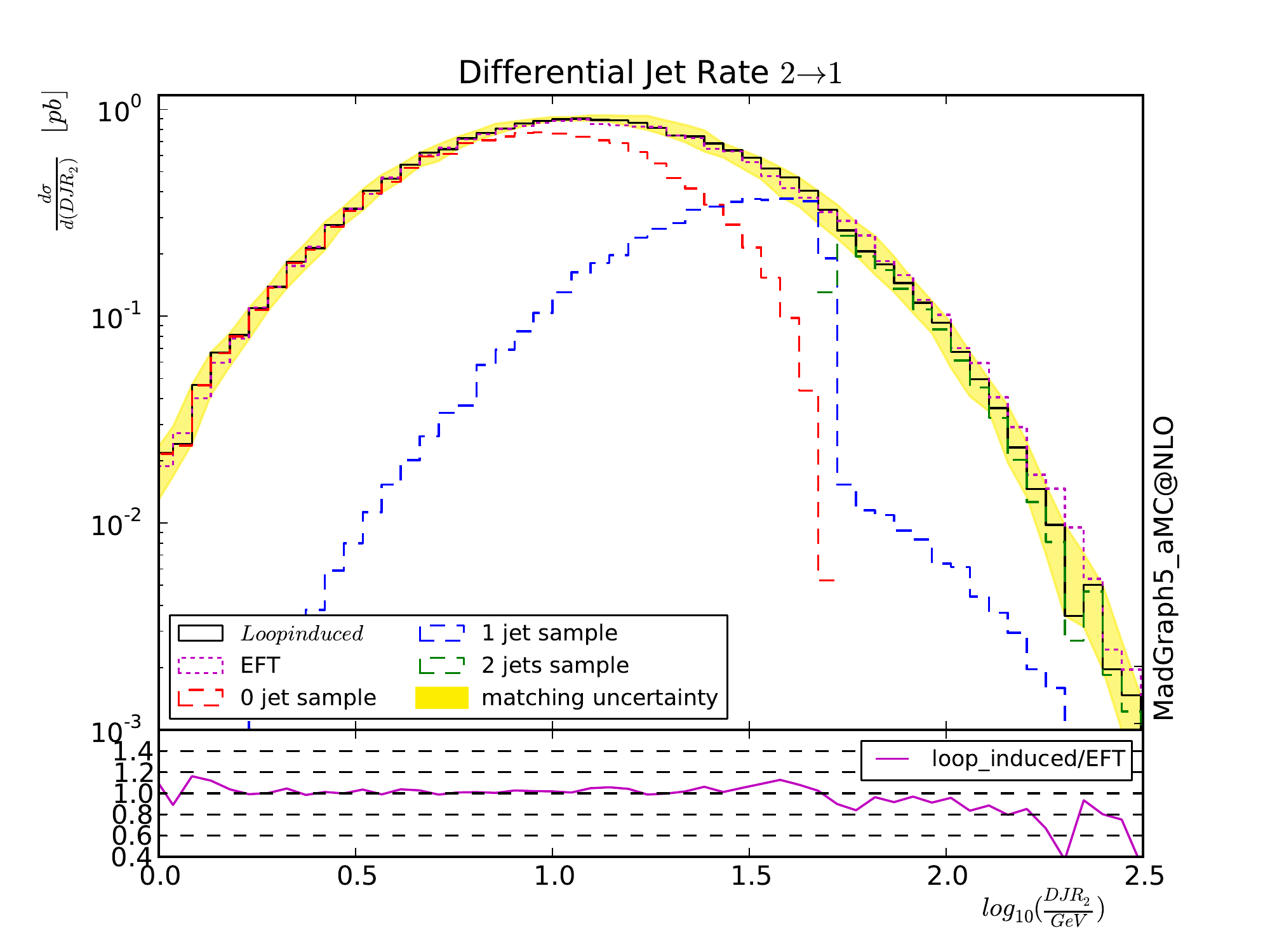}
	\caption{\label{fig:DJR} First two Differential Jet Rate (DJR) histograms for exact loop-induced Higgs production predictions matched to \pythia6 and merged with up to two additional jets. The bottom insets show the ratio of the prediction using exact loop-induced matrix elements to the one performed within the Effective Field Theory (EFT) framework.}
\end{figure}

As it is the case for tree-level computations, \mgamc\ can perform matched and merged computations in the context of loop-induced processes. The MLM merging (either $k_T$-MLM  or shower $k_T$~\cite{Alwall:2008qv}) is available and fully automatic when linked to \pythia6~\cite{Sjostrand:2006za}. As an example of such computation, we present plots for Higgs production merged with up to two additional jets in the $k_T$-MLM scheme. In Fig.~\ref{fig:DJR}, we present the $N^{th}$ differential jet rate plot (DJR$_{<N>}$) which corresponds to the scale for which the $k_T$ clustering algorithm switches from an N to N-1 jet description, that is the scale associated to the N${}^{th}$ emission. This observable is most sensitive to the matching/merging since it is directly related to the variable disentangling matrix-element emissions from parton shower ones. The parton shower is not allowed to radiate in the phase-space region above the merging scale (set here to $Q_{match}=50$ GeV as in~\cite{Alwall:2011cy}) since this region is already populated by partonic events generated using matrix elements of higher multiplicity.\footnote{There are two exceptions to this rule: first, for the highest multiplicity sample, the shower can radiate up to the scale of the softest matrix element jet and secondly a jet can be excluded from the matching if an emission is due to a diagram which does not have a counterpart in the parton shower description. In the latter case, the event is seen by the matching procedure as having one less jet, since the corresponding jet emission is excluded. This explains why the dashed blue curve (corresponding to the single jet multiplicity) does not vanish above the matching scale for the DJR$_2$ plot (left panel of fig.~\ref{fig:DJR}).}
Therefore, having a smooth transition at $Q_{match}$ for the sum of samples (distribution in black) is an important check of the quality of the matching/merging procedure.
For DJR$_1$, a small discontinuity can be observed, originating from the fact that the bottom and top quark interference effect is not accounted for by the parton shower. 

\begin{figure}
	\centering
	\includegraphics[trim=0 50 0 30,clip,scale=0.40]{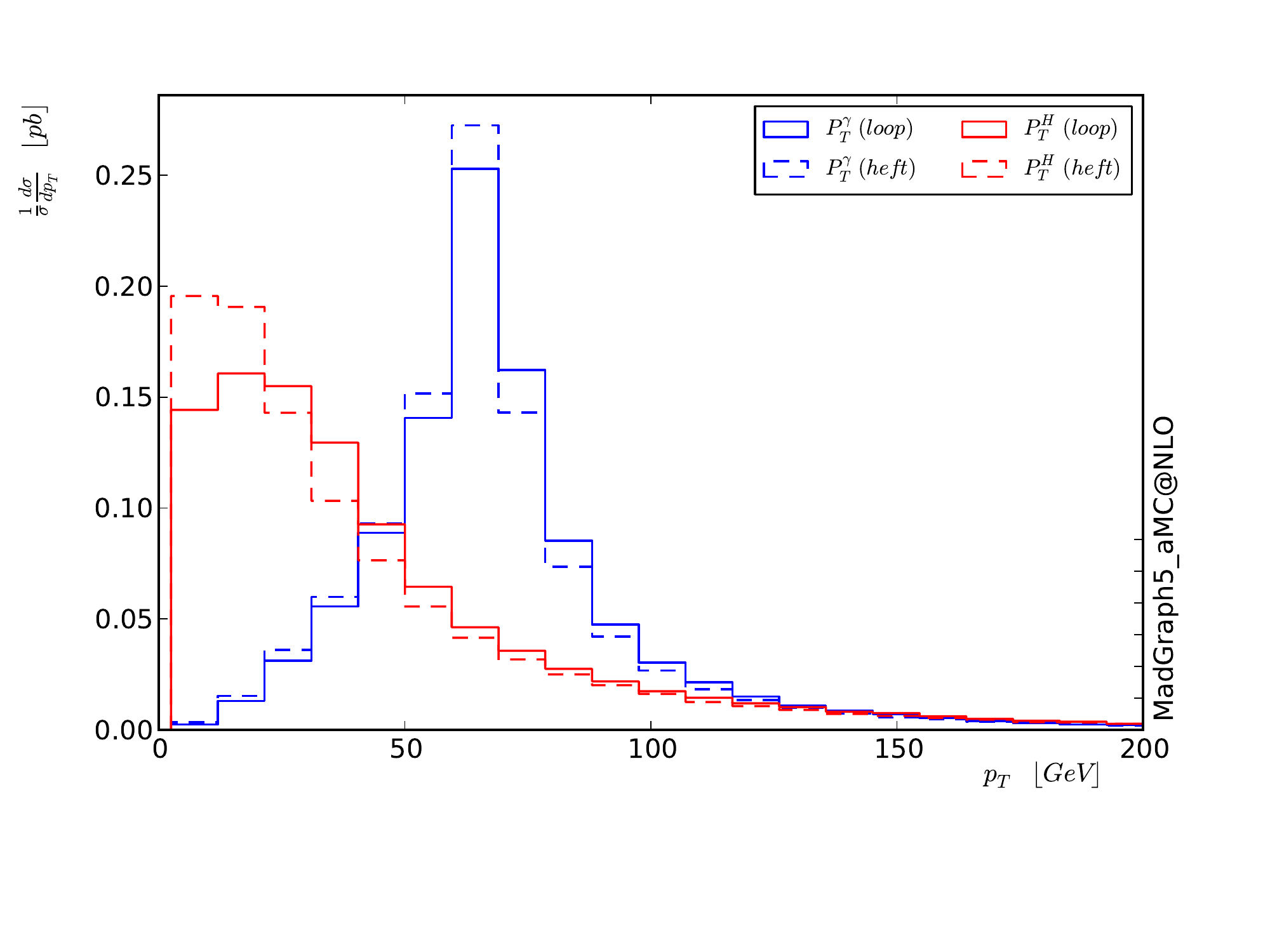}
	\caption{\label{fig:Hdecay} Higgs transverse momentum differential distribution in a matched/merged ($k_T$-MLM, $Q_{match}=50$ GeV) prediction with up to two jets simulated with matrix elements. We present $p_T$ distributions for both the Higgs and the hardest ($p_T$-ordered) photon into which it decays.}
\end{figure}

Finally, fig.~\ref{fig:Hdecay} shows the normalised distribution of the Higgs transverse momentum obtained from the same matched and merged sample. This plot also presents the transverse momentum of the hardest photon when considering the Higgs decaying into two photons (simulated in the narrow-width approximation). Notice that the discrepancy between the EFT and the loop-induced prediction is only due to differences in the production mechanism since the (loop-induced) matrix element for any 2-body decay is a constant over phase-space. 

\section{Loop-induced processes within BSM models: the 2HDM example}
\label{sec:Validation}

Loop-induced processes simulation with \mgamc\ is not limited to the Standard Model but can also be extended to a large class of BSM models. The framework supports models following the NLO UFO format~\cite{Degrande:2011ua,deAquino:2011ub,UFONLO} which can be automatically created by FeynRules/NLOCT~\cite{Alloul:2013bka, Degrande:2014vpa}. Since loop-induced processes are finite, it is not required to have UV counterterms specified in the NLO UFO model, therefore allowing to compute loop-induced processes even for models for which automatic UV renormalisation is currently not possible. The computation of the $R_2$ Feynman rules is however still necessary (see sect.~\ref{techloop}).

\begin{table}[ht]
\begin{center}
 \begin{tabular}{l|c|rr|rrrrr} \hline \hline
 & Type& $\tan\beta$ & $\alpha/\pi$ & $m_{h^0} $ & $m_{H^0} $ &  $m_{A^0} $  & $m_{H^{\pm}} $   & $m^2_{12} $  \\ \hline 
B1 & II & 1.75 & -0.1872 & 125 & 300 & 441 & 442 & 38300  \\
B2 & I & 1.20 & -0.1760 & 125 & 200 & 500 & 500 & -60000 \\
B3 &II & 1.70 & -0.1757 & 125 & 350 & 250 & 350 & 12000  \\ \hline \hline
 \end{tabular}
 \end{center}
 \caption{ Benchmark parameter sets for the two Higgs Doublet Model (2HDM) used for the computations presented in this section. All masses are given in GeV. The top quark mass is set to 173 GeV and the bottom quark mass to 4.75 GeV.
 }
 \label{tab:benchmarks}
\end{table}

In this section, we present results in the two Higgs Doublet Model (2HDM)~\cite{Degrande:2014qga} for the three benchmark parameter sets presented in table~\ref{tab:benchmarks} and introduced in refs.~\cite{Hespel:2014sla, Hespel:2015zea}. Ref.~\cite{Hespel:2015zea} computes the cross-section of the associated production of a Z boson with all the 2HDM scalars using a reweighting approach. As a validation of our method, we have reproduced the computation presented in table 6 of ref.~\cite{Hespel:2015zea}. The only two differences with respect to the parameters used in sect.~\ref{sec:results} is the c.o.m. energy set to 14 TeV and the renormalisation and factorisation scales set to the partonic invariant mass. Our results  are reported in table~\ref{tab:2HDM_cross} and show perfect agreement.

\begin{table*}
\renewcommand{\arraystretch}{1.3}
\begin{center}
    \begin{tabular}{ l | c | c | c }
        \hline \hline
       & $gg\to Zh^0$
         & $gg\to ZH^0$
         & $gg \to ZA^0$
         \\
         \hline  
B1&
 113.6 $^{+28.9\%}_{-21.2\%}\,\,^{+1.0\%}_{-1.2\%}$&
 682.4 $^{+29.6\%}_{-21.5\%}\,\,^{+1.2\%}_{-1.2\%}$&
 0.6203 $^{+32.5\%}_{-23.0\%}\,\,^{+1.9\%}_{-1.9\%}$\\
B2&
 85.59 $^{+29.9\%}_{-21.4\%}\,\,^{+1.4\%}_{-1.1\%}$&
  1545 $^{+30.1\%}_{-21.8\%}\,\,^{+1.3\%}_{-1.3\%}$&
 0.8614 $^{+33.0\%}_{-23.3\%}\,\,^{+2.0\%}_{-2.0\%}$\\
B3&
 169.9 $^{+28.1\%}_{-19.9\%}\,\,^{+1.4\%}_{-0.5\%}$&
 0.8968 $^{+31.2\%}_{-22.3\%}\,\,^{+1.5\%}_{-1.6\%}$&
  1317 $^{+28.4\%}_{-20.8\%}\,\,^{+1.0\%}_{-1.0\%}$\\         
\hline \hline
\end{tabular}
 \caption{\label{tab:2HDM_cross} Inclusive cross sections (in fb) for loop-induced Z boson production in association with each neutral Higgs of the 2HDM model at the 14 TeV LHC. We report results for the three benchmark parameters sets introduced in table~\ref{tab:benchmarks}. The first uncertainty (in percent) refers to scale variations by factors  $\frac{1}{2}$,1 and 2 while the second one refers to PDF uncertainty. Both are computed by \syscalc\ at no additional computational cost.
 }  
\end{center} 
\end{table*}

\begin{table*}
\renewcommand{\arraystretch}{1.3}
\begin{center}
    \begin{tabular}{ l | c | c }
        \hline \hline
       & $gg\to H^+ H^-$
         & $q\bar q \to H^+ H^-$
         \\
         \hline  
B1&
 0.2334 $^{+34.0\%}_{-23.8\%}\,\,^{+2.2\%}_{-2.2\%}$&
 0.7669 $^{+5.9\%}_{-5.4\%}\,\,^{+1.1\%}_{-1.0\%}$\\
B2&
 0.7011 $^{+34.6\%}_{-24.1\%}\,\,^{+2.4\%}_{-2.4\%}$&
 0.4406 $^{+6.5\%}_{-5.9\%}\,\,^{+1.4\%}_{-1.0\%}$\\
B3&
 0.618 $^{+32.8\%}_{-23.2\%}\,\,^{+1.9\%}_{-1.9\%}$&
 2.072 $^{+4.6\%}_{-4.3\%}\,\,^{+0.9\%}_{-0.8\%}$\\    
\hline \hline
\end{tabular}
 \caption{\label{tab:2HDM_cross_charged} Inclusive cross sections (in fb) for the production of a pair of 2HDM charged scalars at the 14 TeV LHC for the three benchmark parameters sets introduced in table~\ref{tab:benchmarks}. The first uncertainty (in percent) refers to scale variations by factors  $\frac{1}{2}$,1 and 2 while the second one refers to PDF uncertainty. Both are computed by \syscalc\ at no additional computational cost.
 }
\end{center} 
\end{table*}

\begin{figure}
	\centering
	\includegraphics[clip,scale=0.25]{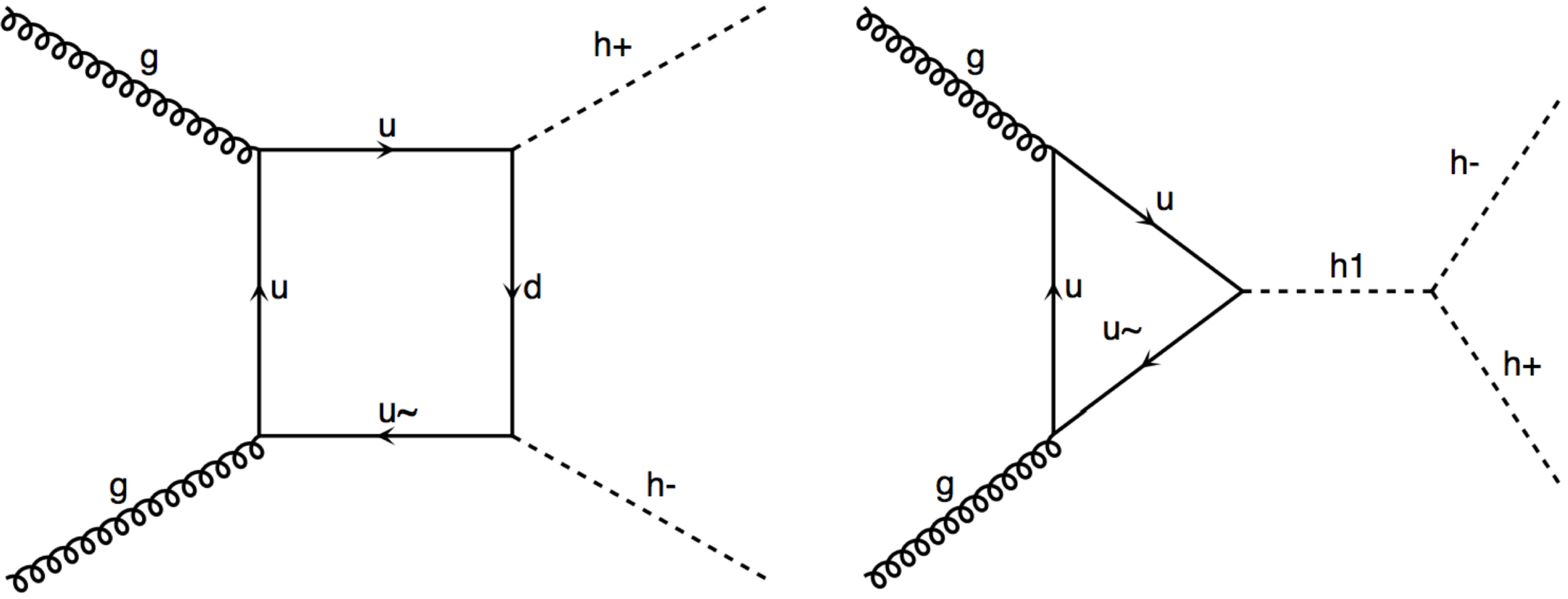}
	\caption{\label{fig:chargedHiggs} Representative diagrams for the production via gluon fusion of a pair of charged Higgs in the two Higgs doublet model.}
\end{figure}

\begin{figure}
	\centering
	\includegraphics[trim=10 0 40 20,clip,scale=0.37]{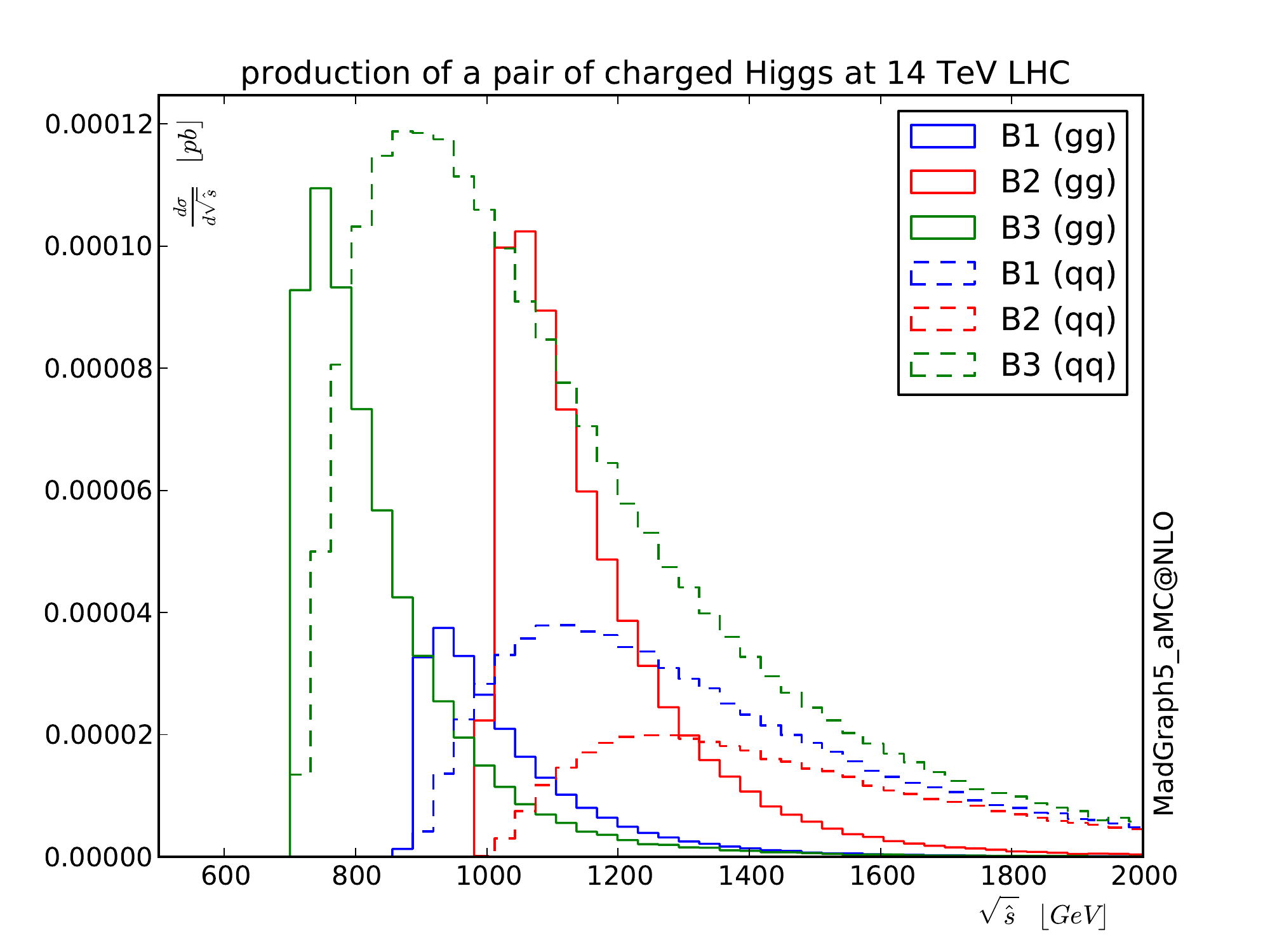}
	\includegraphics[trim=10 0 40 20,clip,scale=0.37]{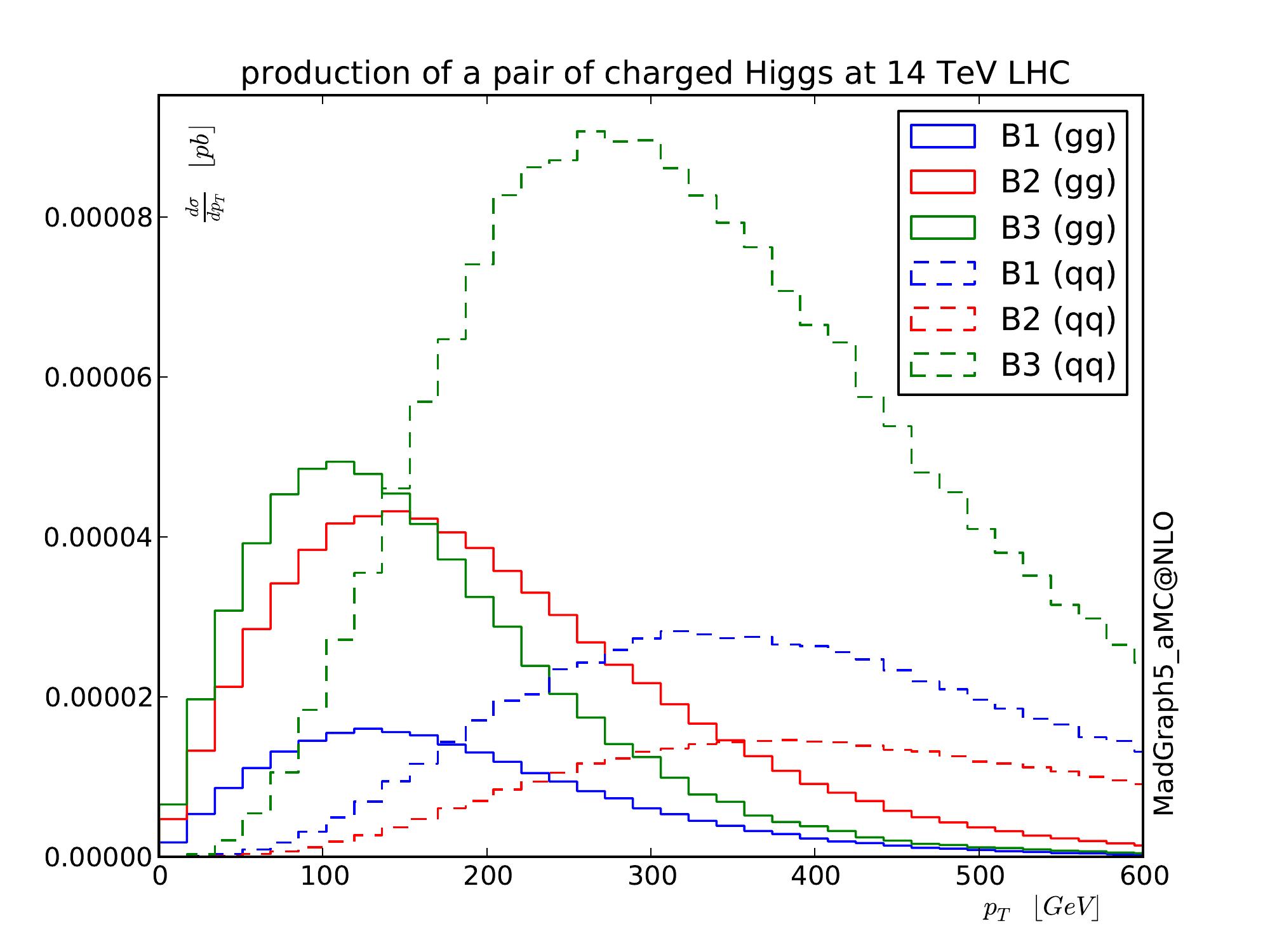}
	\caption{\label{fig:Higgs_charged} Differential distributions at the 14 TeV LHC for the production of a pair of charged Higgs. We present separately the loop-induced gluon initiated production (solid line) and the tree-level quark initiated one (dashed line). 
	}
\end{figure}

In table~\ref{tab:2HDM_cross_charged}, we present  results for the production of a pair of charged Higgs via gluon fusion (see Fig.~\ref{fig:chargedHiggs}) and we compare it to the tree-level channel with initial-state quarks.
Even if the loop-induced contribution is here formally NNLO, it is numerically important. Moreover, the loop-induced contribution differs significantly in its kinematic distribution, as illustrated in Fig.~\ref{fig:Higgs_charged} where we show the system invariant mass and the transverse momentum distribution of each partonic subprocess. For all benchmark points, the distributions are softer for the loop-induced production channel because of the differences in the relevant parton density functions.

\section{Conclusion}
\label{conclusion}

We have presented a new public extension of \madgraphamc\ capable of automatically simulating loop-induced processes within both the Standard Model and BSM theories.
Our implementation does not rely on any predefined set of processes and instead generates on demand an optimised code tailored to the loop-induced process of interest. This is achieved by combining the capabilities of \MadLoop\ for the generation of loop-induced matrix elements with those of \madevent\ for phase-space integration.

Particular emphasis is put on run time and we have optimised both aforementioned codes for the purpose of loop-induced processes whose matrix-elements are considerably slower to evaluate than tree-level ones. We have first determined that, within our framework, the most efficient approach is to employ the OPP reduction method for a given helicity configuration, which we pick for each phase-space point according to a probability set by importance sampling. A second improvement is the computation of the loop matrix elements by projecting its colour structures onto the colour-flow basis. This also provides the associated leading colour information necessary for event generation and required by parton shower programs.
Finally, we have restructured \madevent\ code so as to allow for an arbitrary parallelisation of the computation.

Thanks to these developments, it is possible to compute all $2\to2$ loop-induced processes on a laptop and all $2\to3$ ones (and most $2\to4$) on a small size cluster. As a proof of this statement, we have computed a large list of cross-sections in the Standard Model including all loop-induced processes up to $2\to3$ and a couple of $2\to4$ processes as well as some decay and non-hadronic scattering processes. The implementation was carefully validated against results in the literature for both integrated distributions and local phase-space points.

Our approach is fully differential and compatible with standard matching and merging procedures like the MLM approach supported by \mgamc; this is emphasised by phenomenological results and distributions presented for the case of Higgs production. Finally, the flexibility and generality of \mgamc\ regarding the support of BSM models is maintained and we illustrated this with results for various loop-induced processes within the two Higgs doublet model.

This work fills a gap in the spectrum of modern tools capabilities and it opens the door to many applications whose exploration will no longer be hindered by the technical difficulties of simulating loop-induced processes.

\acknowledgments

We would like to thank all the authors of \mgamc, and in particular Fabio Maltoni and Tim Stelzer, for their help and support at many stages of this project.
We would also like to  thank Benoit Hespel, Stefan Hoeche, Huasheng Shao and Eleni Vryonidou,  for their help during the validation of our implementation.
Finally, we thank the CP3 IT team for their constant support.
V.H. is supported by the Swiss National Fund for Science (SNFS) with grant P300P2 161050.
O.M. is supported by a Durham International Junior Research Fellowship.
This work is supported in part by the IISN  "\MadGraph" convention 4.4511.10, by the Belgian
Federal Science Policy Office through the Interuniversity Attraction Pole P7/37, by the European Union as part of the FP7 Marie Curie Initial Training Network MCnetITN (PITN-GA-2012-315877), and by the ERC grant 291377
`LHCtheory: Theoretical predictions and analyses of LHC physics:
advancing the precision frontier'.

\appendix
\newpage
\section{Manual}
 \label{app:manual} 
In this appendix, we present minimal information on how to generate events for a given loop-induced process. We then proceed with the description of the new options introduced in \mgamc~version 2.3.0 and their particular application in the context of loop-induced, tree-level and NLO computations.

\subsection{Basic commands}
\label{basiccommands}
The steps needed to generate loop-induced processes are very similar to the ones needed to generate a tree-level and/or an NLO computation within the \mgamc\ framework. For the reader familiar with that environment, it should be enough to stress that for processes without any Born diagrams, the `{\tt generate}' command suffixed with the NLO syntax (\emph{i.e.} adding `{\tt{[}QCD{]}}' at the end of the process definition) will automatically consider the corresponding loop-induced process.  From now on, we adopt a pedantic approach and do not assume any prior knowledge of the tool.

The \mgamc\ framework is a meta-code generating optimised numerical programs for the simulation of user-defined scattering/decay processes~\cite{Alwall:2014hca}. The code offers an interactive interface, that can be started with the script {\tt{<MG\_install\_path>/bin/mg5\_aMC}}, where the user can enter commands to specify the processes/runs he is interested in.\footnote{All those steps can also be done without using the interactive interface (\emph{i.e.} scripting). For more information on this please read: https://answers.launchpad.net/mg5amcnlo/+faq/2186} All commands and options are documented within the code, and detailed information can be obtained by typing `{\tt help COMMAND}'. The list of all commands available are listed by typing `{\tt help}'. Finally, a built-in tutorial can be started using the command `{\tt tutorial}'.
We now proceed to describing the succession of the four commands characteristic of the steering of a simulation.

\begin{itemize}
\item {\bf import model \emph{<MODEL\_NAME>}:}\\
This command allows to import a new model. 
In the context of loop induced processes, only NLO \ufo\ models are supported. Such models can be generated via \feynrules~\cite{Alloul:2013bka} thanks to the NLOCT package~\cite{Degrande:2014vpa}. If this command is skipped, a simplified SM model (with diagonal-CKM and massless quark/leptons up to second generation) is loaded by default.\\
\underline{examples:}\\
\hspace*{.7cm}import model loop\_qcd\_qed\_sm\\
\hspace*{.7cm}import model 2HDM\_NLO\\

\item {\bf generate \emph{<PROCESS\_DEFINITION>}:}\\
This line corresponds to the definition of the process of interest and to the generation of the corresponding Feynman diagram. The process definition consists in a string with the initial state and final state particles (separated by `{\tt >}'). To specify a loop-induced process, one must suffix the process definition with the tag `{\tt {[}QCD{]}}' (same as for NLO QCD computations). In the presence of this tag, \MadLoop\ first checks whether it receives tree level contributions; if it does, then a standard NLO computation is performed. It is possible to use the tag `{\tt {[}noborn=QCD{]}}', forcing \MadLoop\ to build matrix elements using the square of loop diagrams (\emph{i.e.} loop-induced mode) without checking for the existence of tree-level contributions.
\\
\underline{examples:}\\
\hspace*{.7cm}generate p p > h j {[}QCD{]}\\
\hspace*{.7cm}generate p p > t t {[}QED{]}\\
\\
If one is interested in only the loop matrix element evaluation for a given kinematic configuration (\emph{i.e.} not including the integration and event generation related code), the tag {[}virt=QCD{]} should be used instead. The subsequent `{\tt launch}' command then asks for the specification of the phase-space point and model parameters and returns the corresponding numerical value of the loop matrix-element. A dynamic library of the fortran code produced by \MadLoop\ for the matrix element computation can then easily be produced by running the shell command `{\tt make OLP}' in the `\emph{SubProcesses}' directory of the process output folder.
\\
\underline{example:}\\
\hspace*{0.7cm}generate h > g g {[}virt=QCD{]}

\item {\bf output \emph{<PATH>}:}\\
This starts the generation, in the specified path, of the numerical code for the process of interest.
\\
\underline{example:}\\
\hspace*{.7cm}output MY\_NEW\_RUN

\item {\bf launch \emph{<PATH>}:}\\ 
This command starts the actual computation of the cross-section and the generation of events. 
The code first asks two preliminary questions. First, the list of external tools (\pythia6~\cite{Sjostrand:2006za}, \delphes~\cite{deFavereau:2013fsa}, \madspin~\cite{Artoisenet:2012st}, ...)
that one wishes to use in this simulation can be turned on. Then, the user is given the option to edit various configuration files, including the model parameters.
This part is identical to the case of leading order computations and we refer to~\cite{Alwall:2014hca} for more details in this subject. Once the simulation is completed, various outputs are written in the folder `\emph{<PATH>/Events}' and a convenient html summary is generated in `{\emph{<PATH>/crossx.html}}'. 

\underline{example:}\\ 
\hspace*{.7cm}launch \emph{MY\_NEW\_RUN}

\end{itemize}

\subsection{Description of new \madgraphamc\ options}

\label{subsec:newoptions}
In this new version of \mgamc\ (version 2.3.0), we have introduced a couple of additional options related to the simulation of loop-induced processes.
The following options can be modified in the file `\emph{input/mg5\_configuration.txt}' or directly via the interface using the `{\tt set}' command.

\begin{itemize}
\item {\tt cluster\_size} [\emph{default=100}]: This parameter allows to arbitrarily scale the requested number of jobs to be proportional to the size of the cluster. By increasing the cluster size, \mgamc\ splits iterations into more jobs, each probing less phase-space points. As of now, this parameter applies for the simulation of loop-induced processes only.

\item {\tt cluster\_local\_path} [\emph{default=None}]: This parameter avoids either to transfer PDF sets to the cluster nodes or to read them directly on a central disk.
This path should point to a (node specific) directory containing the associated PDF sets (either those from LHAPDF or built-in ones). A typical usage is to set this path to a local directory mirrored via {\sc\small cvmfs}.

\item {\tt max\_npoint\_for\_channel} [\emph{default=3}]: This parameter controls what topologies enter the multi-channeling to be used for integrating loop-induced processes. For instance, when set to 4, all loop diagrams with 4 or less loop propagators seed their own channel of integration. This means that tree topologies obtained by shrinking \emph{up to box} loop diagrams are considered for the multi-channeling and therefore integrated separately. In general, we do not observe any significant gain (when not detrimental) when setting this parameter to 4 or larger (except in the case of $g g \rightarrow z z$). We stress here that, for this parameter to take effect, it must to be modified prior the generation of the source code of the process considered.

\item {\tt loop\_colour\_flows} [\emph{default=dynamical}]: The computation of partial colour subamplitudes (\emph{i.e.} amplitudes for fixed colour flows) is turned off by default for the case of NLO virtual matrix elements and turned on for loop-induced matrix elements.. This is because in the former case it comes at the price of giving up loop reduction at the squared amplitude level, hence slowing down \MadLoop\ execution speed since the number of OPP reductions is no longer independent of the number of contributing helicity configurations. This option can however be turned on (before \MadLoop\ writes out the source code for the process) since colour subamplitudes can be necessary for certain applications, such as NLO event generation within the context of a controlled colour expansion~\cite{Frixione:2011kh} and/or Monte-Carlo over colours. Partial colour amplitudes (called {\tt JAMP} in the code) can then be accesses and combined as needed by modifying the user-defined subroutine `\emph{*\_COMPUTE\_COLOR\_FLOWS\_DERIVED\_QUANTITIES}' present in the source code file `\emph{compute\_color\_flows.f}'.

\item {\tt loop\_optimized\_output} [\emph{default=True}]: This option corresponds to using the polynomial decomposition of the integrand numerator (Eq. \ref{eq:openloop}), introduced in~\cite{Cascioli:2011va}, to optimise the use of OPP reduction and allow to interface TIR tools. For debugging and validation purposes it is useful to be able to turn this option off (before generating the loop matrix element code) and to force \MadLoop\ to recompute the complete integrand numerator for each new value of the loop-momentum specified by the OPP reduction procedure. This provides a strong check on the correctness of the computation of the polynomial coefficients.

\end{itemize}

On top of the parameters above which control the way the code is generated and then handled by the cluster, we also introduced new entries in `\emph{run\_card.dat}', which can be accessed and modified at any time, including after code generation.

\begin{table*}
\renewcommand{\arraystretch}{1.3}
\begin{center}
    \begin{tabular}{ c | l c}
    \hline
    \hline
    value & {\tt dynamical\_scale\_choice} meaning\\
    \hline
    -1 & default case:\\
        & \multicolumn{2}{l}{\hspace*{10pt}LO code: transverse mass of the 2 $\to$ 2  system resulting}\\
        & \hspace*{25pt}  of a $k_T$ clustering \\
        & \hspace*{10pt}NLO code: sum of the transverse mass divide by 2 & $\frac12\sum_{i=1}^{N} \sqrt{ m_i^2+p_{T,i}^2 }$.\\
    0  & user defined scale, specified in the file \emph{'setscales.f'} \\
    1 & total transverse energy of the event & $\sum_{i=1}^{N}\frac{E_i\cdot p_{T,i}}{\sqrt{p_{x,i}^2+p_{y,i}^2+p_{z,i}^2}}$.\\
    2 & sum of the transverse mass &  $\sum_{i=1}^{N} \sqrt{m_i^2+p_{T,i}^2 }$.\\
    3 & sum of the transverse mass divide by 2 & $\frac12\sum_{i=1}^{N} \sqrt{m_i^2+p_{T,i}^2 }$.\\
    4 & partonic energy & $\sqrt{\hat s}$.\\
    \hline
    \hline
\end{tabular}
 \caption{\label{table:dynamical_scale_choice} Values supported in `\emph{run\_card.dat}' for the new entry {\tt`dynamical\_scale\_choice}' and their corresponding functional form of the factorisation and renormalisation scale $\mu_F$ and $\mu_R$. This parameter applies for both leading order and next to leading order computations.}  
\end{center} 
\end{table*}

\begin{itemize}
\item {\tt nhel}: This parameter was already defined in previous versions of the code but its effect changed in \mgamc\ v2.3. `{\tt nhel}' can now only be set to `0' or `1', in which case all, respectively exactly one, helicity configuration(s) are/is considered for each phase-space point. The difference with previous versions is that the helicity configuration picked for each phase-space point is chosen according to a dynamical importance sampling Monte-Carlo method. The default value for this parameter is taken to be `1' for loop-induced process and `0' for tree level computations (this parameter is not available for NLO computations). We stress that when setting this parameter to `0' for loop-induced processes, it is more efficient to select the tensor integral method {\sc\small PJFry} or {\sc\small IREGI} as the preferred reduction by modifying the parameter `{\tt MLReductionLib}' of the `\emph{MadLoop\_card.dat}' configuration file.

\item {\tt dynamical\_scale\_choice}: When a dynamical scale is chosen, this parameter selects its functional form (\emph{c.f.} table~\ref{table:dynamical_scale_choice} for a list of predefined functional forms).

\item {\tt survey\_splitting}: For the phase-space integration of loop-induced processes, this parameter allows to define the number of nodes assigned to the integration of one given channel integration during the survey/gridpack generation. Notice that the parallelisation of the second stage of the simulation (which completes event generation and is referred to as the \emph{refine} step) is controlled by the parameter {\tt cluster\_size} instead.

\item {\tt job\_strategy}: This parameter controls the parallelisation strategy for the simulation of high multiplicity tree level simulations (especially important for an efficient handling of the simulation of the highest multiplicity sample in the context of an MLM merging gridpack generation.). The three possible integer values taken by this option are:
\begin{itemize}
\item `{\tt 0}': The original behaviour already adopted in previous versions where each submitted jobs performs successively the integration of two channels of integration.
\item `{\tt 1}': This mode changes the behaviour for the highest multiplicity sample, where the associated jobs run a single channel of integration. The number of jobs submitted in this case is therefore increased by two, hence reducing the running time for this sample by the same factor.
\item `{\tt 2}': This mode uses the same algorithm as the one introduced for loop-induced computations for the highest multiplicity sample and uses  the mode `{\tt 1}' for the next-to-highest multiplicity sample.
 
\end{itemize}

\end{itemize}

The default value of some of the `\emph{run\_card}' parameters are now dynamically chosen depending on the process considered, so as to better reflect what is typically expected in this case. 
These parameters remain accessible and can be modified by the user; only their default value is changed.
The parameters for which the default value can differ depending on the process considered are:
\begin{itemize}
\item {\tt energy of the beam}: for electron-positron collision, the default energy is set to 1 TeV (500GeV per beam). The default collision energy for all other processes is $13$ TeV.
\item{\tt PDF type}: if the particles in the initial states of the process are constituted of quarks, gluons or photons then the proton PDFs are used. Otherwise, this parameter is set to the fixed energy scheme without PDF.
\item{\tt MonteCarlo over helicity}: Monte-Carlo over helicity is turned off for tree-level simulations and turned on for loop-induced ones. This options is not available for NLO computations.
\item {\tt maxjetflavour/asrwgtflavour}: set accordingly to the number of quark flavours appearing in the initial states.
\item {\tt cuts}: for (LO) width computation, all cuts are turned off by default. Unlike previous versions of the code, the cuts defined in the {\tt run\_card.dat} are now applied to the computation of the partial width as well.
\item {\tt matching parameter}: At LO, if all the processes considered only differ by their jet multiplicity, the MLM matching/merging (ickkw=1) scheme is turned on by default with a 
parton level cut (`{\tt xqcut}') of 30 GeV.
\end{itemize}

Finally, a new \madspin\ option allows to simulate arbitrary decays (including three-body and loop-induced decays) but without any spin correlation (between production and decay) and without BreitWigner smearing. The detailed description of this new functionality is deferred to a future work, and here we limit ourselves to mentioning that this feature can be used by adding the line `{\tt set spinmode none}' at the beginning of the \madspin\ configuration file.

\section{Benchmark results for various loop-induced Higgs production processes}
\label{localRes}
We start here by presenting the numerical result for the evaluation of the matrix element for the loop-induced process $g g \rightarrow h h g g$, chosen both for its complexity and importance as a background to double Higgs production in vector boson fusion. The SM parameters used in this computation are those specified in table~\ref{tableParams} with $\alpha_s$ set to $0.118$. The kinematic configuration considered is:
\begin{footnotesize}
  \begin{align*}
    [GeV]         &\phantom{=\textrm{( 0}}E&&           \phantom{\textrm{,0}}p_x                            &&\phantom{\textrm{,0}}p_y                           &&\phantom{\textrm{,0}}p_z                           &&\\    
    p_{g_1}            &=\textrm{( 500}&&                           \textrm{, \phantom{-}0}                               &&\textrm{, \phantom{-}0}                               &&\textrm{, \phantom{-}500}                           &&\textrm{)}\\    
    p_{g_2}            &=\textrm{( 500}&&                           \textrm{, \phantom{-}0}                               &&\textrm{, \phantom{-}0}                               &&\textrm{, -500}                                            &&\textrm{)}\\
    p_{h_3}            &=\textrm{( 148.556611322403}&&\textrm{, -20.0350647739655}&&\textrm{, \phantom{-}36.3342710976818}&&\textrm{, -68.7203295313605}                  &&\textrm{)}\\    
    p_{h_4}            &=\textrm{( 322.824972379014}&&\textrm{, -94.1433807610584}&&\textrm{, -273.713728753781}&&\textrm{, \phantom{-}69.3453772222178}                  &&\textrm{)}\\
    p_{g_5}            &=\textrm{( 138.118065647892}&&\textrm{, -95.9848700486689}&&\textrm{, -88.5772755257372}&&\textrm{, \phantom{-}44.9173801606153}                  &&\textrm{)}\\   
    p_{g_6}            &=\textrm{( 390.500350650690}&&\textrm{, \phantom{-}210.163315583693}&&\textrm{, \phantom{-}325.956733181836}&&\textrm{, -45.5424278514726}                  &&\textrm{)}
  \end{align*}
\end{footnotesize}
We report here the squared loop-induced matrix element\footnote{Notice that the customary one-loop matrix element prefactor $\frac{(4\pi)^\epsilon}{\Gamma(1-\epsilon)}\left ( \frac{\mu_F^2}{Q^2} \right )^\epsilon$ is irrelevant in the context of a finite loop matrix element. The dependence on the renormalisation scheme/scale only comes through the running of the $\alpha_s$.} computed for the phase-space point above, summed over all helicity and colour configurations.
\begin{center}
\begin{tabular}{cc}\midrule\midrule
  $g g \rightarrow h h g g$  & [$\text{GeV}^{-4}$]
  \\\midrule
$|\mathcal{M}^{(LI)}|^2$ & \texttt{1.3033633142042775e-12}
\\\midrule\midrule
\end{tabular}
\end{center}
This particular computation is performed in quadruple precision and stability tests show that all 17 double precision digits are numerically stable. As for NLO computations, the standalone \MadLoop\ output can be used for crosschecking an independent implementation of the calculation (see how in sect.~\ref{basiccommands}). 

We now turn to listing \MadLoop\ performances in table~\ref{tab:MLperf}. On top of the process $g g \rightarrow h g g g$, we also consider $g g \rightarrow h h$, $g g \rightarrow h h g$ and $g g \rightarrow h g g g$ so as to reflect the scaling of \MadLoop\ timings with the multiplicity and number of Feynman diagrams.
\begin{table}
\begin{center}
\begin{tabular}{l|cccc}\midrule\midrule
   & $g g \rightarrow h h $ & $g g \rightarrow h h g$ & $g g \rightarrow h h g g$ & $g g \rightarrow h g g g$
  \\\midrule
\# loop Feynman diag. & 16 & 108 & 952 & 2040 
\\
\# topologies & 8 & 54 & 380 & 540
\\
\# helicity config. & 2 & 8 & 16 & 32 
\\
Generation time & 8.7s & 21s & 269s & 1h36m
\\
Output code size & 0.5 Mb & 0.7 Mb & 1.8 Mb & 3.2 Mb
\\
Runtime RAM usage & 4.7 Mb & 20.5 Mb & 102 Mb & 240 Mb 
\\\midrule
\multicolumn{4}{l}{Timing for the computation of a single helicity configuration}
\\\midrule
OPP as in {\sc\small CutTools} & 2.6ms (19\%) & 40.7ms (16\%) & 795ms (13\%) & 1.03s (15\%)
\\
TIR as in {\sc\small IREGI} & 17.5ms (3\%) & 1.14s (0.6\%) & 65s (0.17\%) & 79s (0.18\%)
\\
TIR as in {\sc\small PJFry} & 3.2ms (15\%) & 190ms (4\%) & 28s (0.38\%) & 29s (0.50\%)
\\
TIR as in {\sc\small Golem95} & 15.1ms (3\%) & 615ms (1.2\%) & 16s (0.67\%) & 19s (0.75\%)
\\\midrule
\multicolumn{4}{l}{Timing for the computation summing all helicity configurations}
\\\midrule
OPP as in {\sc\small CutTools} & 5.2ms (18\%) & 328ms (15\%) & 14.7s (19\%) & 33s (14\%)
\\
TIR as in {\sc\small IREGI} & 18.4ms (5\%) & 1.19s (4\%) & 68.2s (2.6\%) & 82.0s (6\%)
\\
TIR as in {\sc\small PJFry} & 3.8ms (25\%) & 243ms (21\%) & 33s (6\%) & 38.2s (14\%)
\\\midrule\midrule
\end{tabular}
\end{center}
\caption{\label{tab:MLperf}Performances of \MLf\ for the computation of various benchmark processes, tested with the gfortran compiler, no optimizations, v4.8.2, on an i7, 2.7 GHz CPU. The percentages in parenthesis correspond to the fraction of time spent for the computation of the polynomial coefficients (the complementary time is entirely spent in the loop reduction). The number of helicity configurations reported includes only those whose contribution is not analytically zero.}
\end{table}
The profiling presented for these processes, as well as for any other, can be automatically reproduced by running the command `{\tt check profile \emph{}<process\_definition>}' from the \mgamc\ interactive interface.
The timings indicated do not include any numerical stability test nor do they account for the fraction of points for which it is necessary to use quadruple precision arithmetics.
The percentages in parenthesis specify the fraction of the running time spent for the computation of the polynomial coefficients of the numerator integrand. The complementary time is entirely spent in the loop reduction algorithm. Table~\ref{tab:MLperf} illustrates various points discussed in the technical sect.~\ref{techloop}, especially the fact that only the OPP loop reduction time scales with the number of helicity computed. We refrained from showing the timing using {\sc\small Golem95} reduction for the case where the matrix element is summed over helicity configurations, because the recycling of tensorial coefficients is not yet implemented for this tool. In general, even though TIR reduction allows for faster timing when summing over helicity configurations, it remains much larger than for the computation of a single helicity configuration, hence our default approach of using OPP reduction in conjunction with a Monte-Carlo over helicity configurations.

The number of topologies refers to the number of different sets of loop propagator denominators present in the computation. In the computation of the virtual matrix element for NLO predictions, the integrand numerators of all loops sharing the same denominator topology can be added together before being reduced, hence greatly diminishing the number of necessary OPP reductions. This grouping of topologies is not applicable for loop-induced computation when using OPP reduction, but TIR can benefit from it since tensor integrals can be recycled across loops sharing the same denominator topology.

It is interesting to note that the reduction time in TIR is almost the same between the processes $g g\rightarrow h h g g$ and $g g\rightarrow h g g g$, even though the latter has twice as many diagrams. This is mainly because the reduction time is dominated by the reduction of topologies with maximal tensorial rank ($r_{max}=6$ here), of which there is the same number in these two processes.

Generation time, output code size and RAM usage show that \MadLoop\ is light-weight and practical for the computation of loop-induced amplitudes with up to at least 5 external legs, and well-suited for cross-checking other codes for more complicated processes.

\begin{table}
\begin{center}
\begin{tabular}{c|ccc}\midrule\midrule
 Reduction tool & Max. $n_\text{loop\_prop.}$ & Max. rank &  Complex masses \\
 \toprule \
{\sc\small CutTools}   & $10^\dagger$ & $n_\text{loop\_prop.}+1$ & yes \\
{\sc\small IREGI}       & $7^\dagger$ & $7^\dagger$ & yes \\
{\sc\small PJFry}       & $5^\star$ & $n_\text{loop\_prop}$ & no \\
 {\sc\small Golem95}  & $6$ & $\max(6, n_\text{loop\_prop.}+1)$ & yes
\\\midrule\midrule
\multicolumn{4}{l}{}\\
\multicolumn{4}{l}{\small{$^\dagger$: This limitation is not intrinsic to the reduction tool, and is only parametrical}}\\
\multicolumn{4}{l}{\small{\phantom{$^\dagger$:ll}so that it is trivial to increase if proven necessary.}}\\
\multicolumn{4}{l}{\small{$^\star$: Pentagons in {\sc\small PJFry} are formally supported but typically too unstable for integration.}}
\end{tabular}
\end{center}
\caption{\label{tab:MLTIR} Limitations of the different reduction methods interfaced to \MadLoop.}
\end{table}
The various reduction methods compared in table~\ref{tab:MLperf} have different ranges of applicability, which we summarise in table~\ref{tab:MLTIR}. \MadLoop\ checks these constraints at runtime for each loop and for each available reduction tool before using the first one applicable. It is possible that a mixture of different reduction methods is used within a single loop matrix element computation. Notice that, because of a certain class of hexagons which are not supported, {\sc\small PJFry} is by default limited to rank-5 loops. We have disabled that limitation for the present benchmark since the timing is unaffected. Also, {\sc\small IREGI} is the only reduction tool capable of reducing loops with tensorial rank exceeding the number of loop propagators by more than one unit. 

\bibliographystyle{JHEP}
\bibliography{Library,Library2}

\end{document}